# Experimental constraints on the rheology, eruption and emplacement dynamics of analog lavas comparable to Mercury's northern volcanic plains

**F. Vetere[1], S. Rossi[1], O. Namur[2,3], D. Morgavi[1], V. Misiti[4], P. Mancinelli[1], M. Petrelli[1], C. Pauselli[1], D. Perugini[1]**

[1] Department of Physics and Geology, University of Perugia, Perugia, Italy.

[2] Institut für Mineralogie, Leibniz Universität Hannover, Hannover, Germany.

[3] Department of Earth and Environmental Sciences, KU Leuven, Leuven, Belgium

[4] Istituto Nazionale di Geofisica e Vulcanologia, Roma, Italy.

Corresponding author: Francesco Vetere (francesco.vetere@unipg.it)

**Key Points:**

- New viscosity data for Mercury northern volcanic plains lavas are presented.
- Mercury lavas show shear thinning behaviour with a decrease of viscosity of ca. 1 log unit as shear rate ($\dot{\gamma}$) varies from 0.1 to 5.0 $s^{-1}$.
- Heat loss during lava flow and emplacement implies that high effusion rates, >10000 $m^3/s$, are required to cover large distances as observed by MESSENGER (NASA).

**Abstract**

We present new viscosity measurements of a synthetic silicate system considered an analogue for the lava erupted on the surface of Mercury. In particular, we focus on the northern volcanic plains (NVP), which correspond to the largest lava flows on Mercury and possibly in the Solar System. High-temperature viscosity measurements were performed at both superliquidus (up to 1736 K) and subliquidus conditions (1569–1502 K) to constrain the viscosity variations as a function of crystallinity (from 0 to 28%) and shear rate (from 0.1 to 5 $s^{-1}$). Melt viscosity shows moderate variations (4 –16 Pa s) in the temperature range 1736–1600 K. Experiments performed below the liquidus temperature show a decreases in viscosity as shear rate increases from 0.1 to 5 $s^{-1}$, resulting in a shear thinning behaviour, with a decrease in viscosity of ca. 1 log unit. The low viscosity of the studied composition may explain the ability of NVP lavas to cover long distances, on the order of hundreds of kilometres in a turbulent flow regime. Using our experimental data we estimate that lava flows with thickness of 1, 5 and 10 m are likely to have velocities of 4.8, 6.5 and 7.2 m/s respectively, on a 5° ground slope. Numerical modelling incorporating both the heat loss of the lavas and its possible crystallization during emplacement allows us to infer that high effusion rates (> 10000 $m^3$/s) are necessary to cover the large distances indicated by satellite data from the MESSENGER spacecraft.

## 1 Introduction

The eccentricity of the orbit of Mercury, in combination with the planet's vicinity to the Sun, is responsible for its very long days (~ 59 terrestrial daytimes) and, locally, extremely high surface temperatures. The daylight temperature at perihelion, estimated on the surface at the equator, is ~700 K, whereas it decreases to ~350 K at 85°N. During the night, the lack of a shielding atmosphere produces a high loss of thermal energy due to radiation and temperature decreases to ~100 K [*Paige et al.*, 1992; *Vasavada et al.*, 1999].

The surface of Mercury is dominated by a secondary volcanic crust, the majority of which formed between 4.2 and 3.5 Ga [*Head et al.*, 2011; *Weider et al.*, 2012; *Denevi et al.*, 2013; *Byrne et al.*, 2016], with minor explosive volcanic activity until ~ 1.0 Ga [*Thomas et al.*, 2014]. Geochemical mapping using the X-Ray Spectrometer (XRS) and Gamma-Ray Spectrometer (GRS) of the MErcury Surface, Space ENvironment, GEochemistry, and Ranging (MESSENGER) spacecraft [*Solomon et al.,* 2001] revealed that the volcanic crust is Mg-rich and Al- and Ca-poor in comparison with terrestrial and lunar crustal material [*Nittler et al.*, 2011; *Weider et al.*, 2012, 2015; *Peplowski et al.*, 2015]. Mercury's crust is also strongly depleted in Fe [*Izenberg et al.*, 2014; *Weider et al.*, 2015]. This is most likely due to extreme partitioning of iron into the core [*Hauck et al.*, 2013] during early differentiation of the planets under highly reducing conditions (IW-3 to IW-7 with IW being the iron- wüstite oxygen fugacity buffer) [*Malavergne et al.*, 2010; *McCubbin et al.*, 2012; *Zolotov et al.*, 2013; *Namur et al.*, 2016a]. The extremely high sulfur contents measured by MESSENGER (1–3 wt.%; [*Weider et al.*, 2015]) can also be explained by differentiation under reducing conditions [*Namur et al.*, 2016a], as sulfur solubility in silicate melts increases with progressively reduced oxygen fugacity conditions [*McCoy et al.,* 1999; *Berthet et al.,* 2009*; Zolotov et al.,* 2013*; Cartier et al.,* 2014; *Namur et al.,* 2016a].

The largest effusive events on Mercury occurred at the highest latitudes of the northern hemisphere and are represented by lavas with the highest $SiO_2$- and $Al_2O_3$-contents and the lowest MgO-contents detected on the planet [*Weider et al*., 2015; Namur et al., 2016b]. These lavas belong to a single smooth plain deposit referred to as the northern volcanic plains (NVP) [*Weider et al*., 2012; *Denevi et al*., 2013], which was dominantly formed between 3.7 and 3.5 Ga 2011; [*Head et al*., 2011; *Weider et al*., 2012; *Ostrach et al*., 2015; *Byrne et al*., 2016] and covers ~ 6% of the surface of the planet [*Denevi et al*., 2013]. A notable characteristic of this geological sector of Mercury is that it contains well-preserved lava flows with a low crater density. Some of these flows can be followed for distances exceeding 100 km [*Byrne et al*., 2013; *Hurwitz et al*., 2013], making them extremely useful to gain a better understanding of the dynamics of magma emplacement on planetary surfaces in general, and on Mercury in particular. Another distinct feature of these lava flows is their very high $Na_2O$ (up to 8 wt.%) [*Peplowski et al*., 2014, 2015], which have been explained by 15–30 % melting of a plagioclase-bearing lherzolitic mantle source [*Namur et al*., 2016b; *Vander Kaaden and McCubbin*, 2016] and low $Al_2O_3$ contents, responsible for the low viscosity of the lavas [*Charlier et al*., 2013; *Sehlke and Whittington*, 2015].

Several authors have investigated the nature of lavas constituting the NVP through morphological and compositional analyses [*Head et al*., 2011; *Ostrach et al*., 2015; *Weider et al*., 2015], mineralogical analysis [*Namur and Charlier*, 2017; *Vander Kaaden et al*., 2017], flow modelling [*Byrne et al*., 2013] and rheological measurements of analog lavas [*Sehlke and Whittington*, 2015]. The origin of such large lava flows and their lateral extent are controversial and may be due to the low-viscosity of the lava [*Stockstill-Cahill et al*., 2012] and/or very high effusion rates [*Head et al*., 2011; *Ostrach et al*., 2015; *Sehlke and Whittington*, 2015; *Namur and Charlier,* 2017]. However, to the best of our knowledge, there is presently no work that takes into consideration the possible variation of heat loss of the lava and its relationship with effusion rates.

In order to better understand the mechanisms of lava emplacement in the NVP, we provide new viscosity measurements using synthetic material with a composition that has been estimated using the most recent XRS and GRS data from MESSENGER [*Weider et al*. 2015; *Peplowski et al*. 2015; **Table 1**]. Viscosity measurements were performed at both superliquidus and subliquidus temperatures, at varied shear rates, and are combined with numerical models in order to propose a hypothesis for the rheological behaviour, eruption and emplacement dynamics of NVP lavas. We propose that the formation of very large lava flows may adequately be explained by a combination of low viscosity (10-20 Pa s) and high effusion rate (> 10000 $m^3/s$). In particular, we show that the effusion rates necessary to produce NVP lava flows are comparable to those observed in large igneous provinces (LIPs) on Earth, which could be consistent with formation of NVP lavas by adiabatic decompression of the mantle source [*Namur et al*., 2016b]. Our work also shows the critical effect of low $Al_2O_3$ and high $Na_2O$ contents on lava rheology and we believe that accurately measuring these elements and their variability across the planet should be a priority target of the BepiColombo mission [*Benkhoff et al*., 2010].

## 2 Starting materials and experimental techniques

The elemental compositions of NVP lavas were obtained from XRS (normalized to Si) and GRS measurements and presented by *Nittler et al*. [2011], *Weider et al*. [2012; 2015] and *Peplowski et al*. [2014, 2015]. They were recalculated by *Namur et al*. [2016b] on an oxide basis. These authors combined individual maps of Mg/Si, Ca/Si, Al/Si and S/Si and only calculated chemical compositions for pixels for which those four ratios were available assuming that the sum of major oxides is 100 wt.%. The main advantage of this method is that Si contents do not need to be arbitrarily fixed [*Vander Kaaden and McCubbin*, 2016]. According to *Peplowski et al.* [2014, 2015], $Na_2O$ is high in NVP lavas. Consequently, *Namur et al*. [2016b] used a Na/Si ratio of 0.20, similar to the average Na/Si ratio of NVP lavas presented by *Peplowski et al*. [2014]. This procedure resulted in the acquisition of a large compositional range for NVP lavas spanning from 55 to 66 wt.% $SiO_2$, 8 to 20 wt.% MgO, 3 to 9 wt.% CaO and 9 to 16 wt.% $Al_2O_3$ [*Peplowski et al*. 2015; *Weider et al*. 2015; *Namur et al*., 2016b].

As reported by *Nittler et al*. [2011] and Weider *et al*. [2015], Mercury's surface contains high abundances of sulfur (1–3 wt.%). Sulfur solubility in silicate melts increases with decreasing oxygen fugacity [*Berthet et al*., 2009; *Cartier et al*., 2014; *McCoy et al,* 1999; *Namur et al*., 2016a]. Oxygen fugacity during mantle melting and volcanic eruptions on Mercury is traditionally considered as being between IW-2 and IW-7 [*McCubbin et al*., 2012; *Zolotov et al*., 2013] although new models show that most lavas were formed at IW-5.4±0.4 [*Namur et al*., 2016a]. Therefore, the potential effect of sulfur on the rheology of Mercurian lavas might need to be considered. However, it was demonstrated that S has a very minor effect on the polymerization of silicate melts and, hence, on their rheology [*Morizet et al., 2015*]. In addition, NVP lavas contain the lowest S contents among Mercurian magmas (0.5 to 2.7 wt.% S with a median value of 1.64 wt.% *Namur et al.,* [2016a]). Therefore, in this study we consider that sulfur plays a minor role in modulating the physical properties of NVP lavas.

According to the above considerations, for this experimental study, we prepared a S-free representative composition of NVP lavas (**Table 1**). We concentrated on lava compositions from the northernmost regions of NVP (> 70° North), which have the lowest $Al_2O_3$-contents [*Weider et al.*, 2015] but also the highest $Na_2O$ contents [*Peplowski et al.*, 2014] and which have not yet been experimentally investigated for viscosity characterization. The synthetic composition was prepared at the Petro-Volcanology Research Group laboratories of the University of Perugia (hereafter PVRG labs). Our composition has a ratio of non-bridging oxygen to tetrahedrally coordinated cations equal to 0.89 reflecting its high degree of depolymerisation. This corresponds well with compositions reported by *Stockstill-Cahill et al.* [2012] and *Vander Kaaden and McCubbin* [2016]. When plotted in a total alkali versus silica (TAS) diagram, our experimental composition lies between trachy-andesite and trachy-dacite fields, similar to the compositions investigated by *Sehlke and Whittington* [2015] and *Vander Kaaden and McCubbin*, [2016].

Five hundred grams of glass were prepared by melting a mixture of oxides and carbonates at 1873 K for 4 hours in a $Pt_{80}Rh_{20}$ crucible in air. Melting was performed in a Nabertherm HT 04/17 $MoSi_2$-heated box furnace (Nabertherm GmbH, Lilienthal, Germany). The melt was poured on to a brass plate to quench. To ensure homogeneity, the quenched melt (glass) was crushed, re-melted and quenched again using the same technique. This technique ensures compositional homogeneity of the glass [*Vetere et al.*, 2015]. Qualitatively, high fluidity was observed during quenching suggesting a low viscosity of the melt.

Viscosity measurements have been performed in a Gero HTRV 70-250/18 high-temperature tube furnace with $MoSi_2$ heating elements (Gero GmbH, Neuhausen, Germany) operating up to 2073 K at room pressure. Thermal ramps can be precisely controlled via computer using the Eurotherm iTools v. 9.57.11 software (Eurotherm, Worthing, West Sussex, UK). Viscosity measurements were performed with a rotational Anton Paar RheolabQC viscometer head at the PVRG labs. This instrument consists of a sample-filled crucible and a rotating measuring spindle that is immersed into the sample. The crucible hosting the silicate melt is made of $Pt_{80}$-$Rh_{20}$ with an inner diameter of 37 mm (outer diameter 40 mm) and height of 70 mm. The spindle is made of $Al_2O_3$ with a circular section of 12.2 mm in diameter and is fixed with a standard collet chuck to the head of the viscometer. The lower end of the $Al_2O_3$ rod is sheathed by a tight-fitting $Pt_{80}$-$Rh_{20}$ foil (0.2 mm thick) in order to prevent any contamination of the silicate melts during the experimental runs. The rotational viscometer allows measurements under controlled shear rate ($\dot{\gamma}$). This allows us to investigate possible shear thinning (an increase of viscosity with decreasing $\dot{\gamma}$) or shear thickening (a decrease of viscosity with decreasing $\dot{\gamma}$) effects. Methods and procedures described by *Dingwell* [1986] and *Ishibashi* [2009] were applied in order to determine melt and melt + crystals viscosities. With this equipment, viscosity can be measured in the range from 0.1 to $10^5$ Pa s [*Hess et al.*, 1996]. The viscometer was calibrated against NIST 717a standard glass, for which the temperature-viscosity relationship is accurately known (https://www.nist.gov). Reproducibility of measurements on the standard glass is on the order of ± 0.03 log units. Since NVP melt viscosity was expected to be low, before running experiments we calibrated the viscometer using a Wacker silicone standard having viscosity of 10 Pas [*Spina et al.*, 2016a, 2016b]. One hundred measurements were performed and the results showed good reproducibility, with average values of 9.7 ± 0.3 (standard deviation) Pa s.

The furnace hosting the experimental charge is equipped with aluminium cooling heads. These are positioned on the top and the bottom openings of the furnace tube in order to prevent overheating of the viscometer head. Cooling is achieved by a continuous flux of cold water (20° C). The furnace can move vertically using two pneumatic cylinders (details in Morgavi *et al.* [2015]). This has two major advantages as it allows us to: (1) carefully prepare the experimental geometry by precisely positioning the outer and inner cylinders outside the furnace; (2) bring the furnace directly to the experimental temperature while the sample is still outside the furnace, preventing the sample from undergoing the entire thermal ramp.

Two cross-mounted Thorlabs single-axis translation stages with a standard micrometer allow the correct positioning of the spindle and the crucible. An alumina ($Al_2O_3$) rod (length=600 mm; FRIATEC Aktiengesellschaft, Mannheim, Germany) is fixed to the lower part of the structure holding the outer cylinder. Temperature was monitored using an in-house built S-type thermocouple ($Pt_{10}Rh_{90}$ vs. Pt) within an $Al_2O_3$ sheath, positioned at the bottom of the crucible. As the rotation of the viscometer prevents the use of thermocouples directly wired to a controller, OMEGA wireless thermocouple transmitters UWTC-Series were employed (OMEGA Engineering, INC., Stamford, Connecticut, USA). Uncertainty on temperature measurements is on the order of 0.5 K.

Prior to viscosity measurements, ca. 70 g of melt was stirred at 1773 K for 2 hours at strain rates ($\dot{\gamma}$) of 5–10 $s^{-1}$. This allowed for the complete removal of possible gas bubbles and the attainment of a compositionally homogeneous melt [*Dowty*, 1980; *Lofgren*, 1983; *Davis and Ihinger*, 1998; *Armienti*, 2008; *Iezzi et al.*, 2008, 2011; *Pupier et al.*, 2008; *Vetere et al.*, 2013a, 2015]. Samples used in the subliquidus temperature experiments were first melted at superliquidus conditions. The temperature was then decreased continuously to the required subliquidus value at a rate of 5 K/min. At the end of the experiments, samples were quenched by moving them into the cooled head of the furnace. The quench rate was of the order of 100 K/min which was sufficient to avoid the formation of quench crystals.

## 3 Analytical methods

Experimental samples were cored out from the outer cylinder after quenching, mounted in epoxy, ground flat and progressively polished using diamond paste for textural and chemical analysis. The composition of phases, i.e. the starting glass (for superliquidus experiments), the glass matrix and crystalline phases of run-products (for subliquidus experiments) were analyzed using a CAMECA SX100 electron microprobe analyser (EMPA) at the University of Hannover (Germany). Analyses were performed with an accelerating voltage of 15 kV. For glasses, we used a beam current of 8 nA and a defocused beam of 10 $\mu$m. Mineral analyses were performed with a beam current of 15 nA and a focused beam (1 $\mu$m). Counting time on peak was 15–20 s (7.5–10 s for background) for each element. For glasses and minerals, we used the following standards for K$\alpha$ X-ray line calibration: albite for Na, orthoclase for K, wollastonite for Si and Ca, $Al_2O_3$ for Al, $TiO_2$ for Ti, MgO for Mg. Raw data were corrected with the CATZAF software and results are reported in Table 1 and Table S1-S2 (supplementary information). A high-resolution scanning electron microscope with field-emission gun (FE-SEM LEO 1525 - ZEISS), installed at the Department of Physics and Geology (University of Perugia), was used to collect back-scattered electron (BSE) images of the experimental charges. Additional details about the analytical conditions of FE-SEM and EPMA, as well as data reduction procedures, are reported in *Vetere et al.* [2015] and *Namur et al.* [2016a].

Abundance and distribution of mineral phases in each experimental charge were determined on BSE images using the Image-ProPlus 6.0 software. This software was also used to calculate length-width aspect ratios (AS) of the crystals by applying an automatic ellipse fitting procedure. When crystals were in contact, aspect ratios were measured manually. Details on the analytical protocol for image analysis are given in previous works [*Iezzi et al.*, 2008, 2011; *Lanzafame et al.*, 2013]. Representative BSE images are reported in **Figure 1**. In particular, the estimation of mineral phase proportions (in area %) was made by linking grey-level values of BSE images with compositions. No stereological correction was applied [*Iezzi et al.*, 2008; *Vetere et al.*, 2010 and 2013a-b]. Magnification used in image acquisition ranged from 150× to 1600× depending on the size, shape and amount of crystalline phases. For each sample, five to ten BSE images, cut perpendicularly to the rotational axis, were collected and analyzed on different parts of the polished section to ensure that results were statistically significant.

## 4. Results

### *4.1 Mercury's superliquidus melt viscosity and modelling*

The liquidus temperature ($T_L$) of our NVP composition was estimated using the alpha-MELTS software package [*Asimow, et al.*, 2004; *Smith and Asimow*, 2005] providing a value for $T_L$ = 1581 K. Experimental data are in good agreement with this estimate; indeed, the experiment performed at 1569 K contained only 2.1 area % of crystals indicating that the liquidus temperature is slightly higher. This is confirmed by the experiment at T=1600 K in which no crystals were detected.

Twenty-seven superliquidus experiments in the temperature range between 1600 and 1736 K were performed in order to determine the dependence of melt viscosity on temperature (**Table 2** and **Figure 2**). In these experiments, melt viscosity ranges from 4.0 Pa s and 16.3 Pa s. Each of the viscosity values presented in Table 2 is an average of 100 to 500 measurements, collected on timescales from 120 to 240 minutes, with torque values measured continuously. The majority of the experiments were performed at a shear rate of 5 $s^{-1}$. We selected this value because of the very low viscosity of the investigated composition. At lower shear rate the instrument's torque limit is reached and measurements are not feasible. Experiments at 1736 K (M3a-b), 1711 K (M7a-b) and 1691 K (M10a-b) were repeated twice and show a high reproducibility (Table 2). The dependence of viscosity upon shear rate ($\dot{\gamma}$) was investigated with a series of 7 experiments performed at 1672 K with $\dot{\gamma}$ from 5 to 10 $s^{-1}$ (samples M16a-g in Table 2). Larger strain rates were not applied in order to prevent the melt to spill out from the crucible due to its high fluidity. No measurable effect of shear rate on viscosity was detected at superliquidus temperature (Table 2).

The obtained viscosity dataset was used to develop an empirical model of viscosity as a function of temperature. The model is based on the Vogel-Fulcher-Tammann (VFT) equation [*Vogel*, 1921]:

$$\log \eta \text{ (Pa s)} = A + B/(T - T_0) \qquad (1)$$

where T is the temperature in Kelvin and A, B, and $T_0$ are fitting parameters, representing the pre-exponential term, the pseudo-activation energy (related to the barrier of potential energy obstructing the structural rearrangement of the liquid), and the VFT temperature, respectively. The VFT approach accounts for the non-Arrhenian temperature dependence of melt viscosity. Data were fitted using a non-linear least-square regression providing the following parameters: A = - 4.30 (Pa s), B = 6244.9 (K) and $T_0$ = 471.2 (K). This relationship reproduces our experimental data with a $r^2$ value of 0.99 (Figure 2). Note that the fitting parameters reported above are only valid for high-temperature viscosity data. In fact, when comparing our VTF parameters with those presented in *Sehlke and Whittington* [2015] for melts considered similar to those erupted on Mercury, we observe a general agreement with terms A and B, but a disagreement with the $T_0$ parameter. This reflects the inability of the above model to reproduce data below the liquidus temperature.

The comparison of viscosity data presented here to those recently published by Sehlke and Whittington [2015] shows a maximum difference of 5.0 Pa s log (η) (Table 1 and **Figure 3**) mainly due to the different chemical composition of the silicate melt used in the experiments. In this respect, an important feature of our composition is the high $Na_2O$ (8.85 wt.%) and low $Al_2O_3$ (8.95 wt.%) contents, in agreement with MESSENGER data for the most evolved lavas of the NVP [*Peplowski et al.*, 2014, 2015]. These compositional characteristics are responsible for the low viscosity of our silicate melt. Previous experimental data (e.g. *Le Losq and Neuville*, 2013; *Vetere et al.*, 2014; *Stabile et al.*, 2016) are in agreement with the lower melt viscosity described here.

*4.2 Mercury's subliquidus viscosity*

The viscous behaviour of the melt below $T_L$ was investigated using 15 experiments in the temperature range of 1569–1502 K at three different shear rates: 0.1, 1.0 and 5.0 $s^{-1}$. Prior to cooling the melt was kept at 1673 K for 2 h in order to erase possible crystal nuclei. The final dwell temperature was reached using a cooling ramp rate of 20 K/min. Experiments were run for up to 24,000 s. In this temperature range, crystals nucleated and grew, increasing from 2 to 28 area %, as temperature decreased (**Table 3** and **Figure 4a-b**).

The relative abundance of mineral phases, crystal sizes, and shapes do not vary significantly at different shear rates. Olivine (pure forsterite) is the liquidus phase in all experiments. Clinopyroxene appears at a temperature of 1520 K and remains stable in the system down to 1502 K. Clinopyroxene does not show significant compositional evolution, varying from $Wo_{39}En_{61}$ at 1520 K to $Wo_{37}En_{62}$ at 1502 K, although a slight increase of sodium content is observed (from 0.25 to 0.95 wt.%) with decreasing temperature (Table S2, supplementary materials). Image analysis estimates of crystal fractions and those obtained by mass balance calculations agree well and are reported in Table 3 and in **Figure 5**.

The compositional evolution of the melt during crystallization is shown in **Figure 6** (see also electronic supplementary material). $SiO_2$ increases from 62.35 to 65.95 wt.% as temperature decreases from liquidus ($T_L$) to 1502 K. Similar trends are observed for $Al_2O_3$, $TiO_2$, and $K_2O$, whereas CaO and MgO decrease. $Na_2O$ shows a more scattered behaviour, presumably due to the combined effect of devolatilization of this element at high temperature and its incorporation into clinopyroxene crystals.

Viscosity vs. time at constant temperature shows a typical S-shape curve, as reported in **Figure 7a-c-e**. On the right panels of the figure, representative BSE images of experimental samples are also shown for three selected samples (M34, M35 and M36; table 3) at 1510, 1520 and 1533 K at shear rate ($\dot{\gamma}$) of 5.0 $s^{-1}$ (**Figure 7b-d-f**).

As shown in Table 3 and Figures 4 and 7, viscosity at $\dot{\gamma}$ = 0.1 $s^{-1}$ varies between 2.72–4.01 Pa s [log (η)] for temperatures ranging from 1533 to 1502 K. As $\dot{\gamma}$ values increase to 1.0 and 5.0 $s^{-1}$, viscosity [log (η)] varies between 1.93–3.04 Pa s (at 1545 K) and 2.25–3.36 Pa s (at 1502 K), i.e. viscosity decreases as shear rate increases. This points to a shear thinning behaviour of the partly crystallized melt. From the curves displayed **in Figure 7**, the time for crystal nucleation and growth can be evaluated at different temperatures [*Vona et al.*, 2011]. Crystal growth appears to be inversely correlated to the shear rate indicating that the higher the applied shear rate, the lower the time needed for crystals to nucleate and grow. For example, at 1510 K the time to reach a constant viscosity value (plateaux in Figure 7a) at $\dot{\gamma}$ = 0.1 $s^{-1}$ is slightly longer than 4.0 hours. At the same temperature, viscosity reaches a constant value after about 2.0 hours at $\dot{\gamma}$ = 5.0 $s^{-1}$. As temperature increases to 1533 K, the time taken to reach a constant viscosity in the partly crystallized system is about 2.0 hours and 1.0 hour, for $\dot{\gamma}$ of 0.1$s^{-1}$ and 5.0 $s^{-1}$, respectively (Figure 7e).

*4.3 Further rheological considerations*

During crystallization, the shear thinning behaviour becomes evident for all experiments, with viscosity decreasing as the shear rate increases (Figure 4). Noteworthy is the fact that the shear-thinning behaviour also arises at low crystal fractions ($\Phi_c$=0.05) and increases at higher crystal contents (Figure 4a). Following the work from *Sehlke and Whittington* [2015], *Sehlke et al.*, [2014] as well as from *Vona et al.*, [2011 and 2013], we calculate the flow index using the linear regression coefficients derived from our experimental data set (**Table 3** and **Figure 8 a** and **b**; see also supporting information files for details).

Flow index values are relatively low compared to those estimated by *Sehlke et al*. [2014] for Hawaiian basalts, as well as from *Sehlke and Wittington* [2015] for Mercury's analog compositions (Fig. 8b). This is presumably due to the large aspect ratios of crystals in our study (up to 14) that can strongly influence flow index values [Mader et al. 2013]. This highlights the highly non-Newtonian behaviour of our partly crystallized analog composition. As an example, at a temperature of 1533 K and crystal content of ca. 9.0 area % (sample M36 in Table 3) the flow index ($n$) is 0.53. Under these conditions, an increase in shear rate from 0.1 to 5 $s^{-1}$ results in a decrease in viscosity by a factor of 6. In a similar way, experiment M33 performed at 1502 K (crystal content of 28 area % with $n$ = 0.42) shows a decrease in viscosity by a factor of 10 (see Table 3 and Figures 4 and Figure 8). A comparison between the rheological behaviour of our composition with data from *Sehlke and Wittington* [2015] is provided in Figure S1 (supplementary materials), where the change in apparent viscosity with crystal fraction is shown. Results indicate a good general agreement between literature data and our experimental results. A slight deviation is observed at the lowest shear rate possibly due to differences in crystal shapes and distributions.

Flow curves for Mercury lavas superimposed on the pahoehoe to `a`a transition threshold diagram, derived from *Sehlke et al.* [2014], are presented in **Figure 9.** Mercury's lavas show similar characteristics, but at slightly lower crystal fractions (0.05) compared to those studied by *Sehlke et al.* [2014]. As the melt crosses the liquidus temperature (1581 K), pseudoplasticity arises (flow index $n < 0.7$), as shown by the fact that the magma lies in the transition threshold zone (TTZ) in Figure 9. It is not surprising that pseudoplastic behaviour is found at very low crystal content. In fact, *Ishibashi and Sato* [2007] detected pseudoplastic behaviour in alkali olivine basalt at $\Phi$ as low as 0.05 (olivine, plagioclase, and spinel). In addition, *Ishibashi* [2009] detected pseudoplastic behaviour in basalt from Mount Fuji between $\Phi$ of 0.06 to 0.13 due to suspended plagioclase crystals. The transition from pahoehoe to `a`a for the analog Mercurian lava studied here begins at a temperature of ~1533 ± 10 K.

In modelling viscosity vs. crystal content, the relative viscosity (defined as $\eta_r = \eta_{eff}/\eta_m$, where $\eta_{eff}$ is the effective viscosity of the suspension with a volume fraction of crystals, and $\eta_m$ is the viscosity of the melt; see supplementary materials for details) is one of the most used parameter. **Figure 10** shows the variation of $\eta_r$ as crystallinity ($\Phi$) increases. Our experimental data can be described by the Einstein-Roscoe equation (see supplementary information) for relatively high $\dot{\gamma}$. Lowering $\dot{\gamma}$ to the value 0.1 $s^{-1}$ results in higher $\eta_r$ (reaching values up to ca. 15) matching results reported by *Vona et al.*, [2011] for crystallinity higher than 20 vol.%.

## 5. Discussion

According to results presented by *Byrne et al.* [2013] on the lava flows forming the NVP, some important constraints emerge: i) NVP lavas can flow over very long distances, on the order of hundreds of kilometres; ii) the channels filled by the lavas appear quite variable in terms of their width, spanning from a few hundreds meters to tens of kilometres; iii) the morphological characteristics of lava flows, observed using high-resolution images of the planet surface, indicate that lavas were emplaced as turbulent flows [*Byrne et al.*, 2013]. These constraints must be taken into account when attempting to shed new light upon the mechanisms that might have contributed to the emission and emplacement of lava flows forming the NVP. In the following discussion, the new experimental data presented in this work are integrated with the above constraints, in order to refine our understanding of the dynamics of NVP lava flows. Notably, in conducting both superliquidus and subliquidus experiments, we attempt to constrain the behaviour of lavas in terms of velocity and distance covered by the flows, as well as the possible effusion rates.

The initial issue to consider is the slope of the terrains on which the lava was emplaced. Although it would be preferable to use pre-eruption topographic data to estimate lava flow velocity, this is not possible at present and we must rely upon present day, post-NVP emplacement, topography [e.g. *Byrne et al.*, 2013]. Hereafter, we consider average slope values in between 0.1°–5°; these are of the same order of magnitude as those used in other studies of lava flows on the surface of Mercury [e.g., *Byrne et al.*, 2013; *Hurwitz et al.*, 2013].

As the emplacement of lavas on Mercury likely occurred in a turbulent regime [*Byrne et al.*, 2013] we can use the approach proposed by *Williams et al.* [2001], which is valid for high Reynolds numbers ($Re$>>2000), to infer the possible velocity of lava flows. This method allows for the estimation of lava flow velocity $u$ *(m/s)*, considering the ground slope $\theta$ (°) and the lava friction coefficient $\lambda$:

$$u = \sqrt{\frac{4gh\sin(\theta)}{\lambda}} \quad (2)$$

where $g$ *(m/s*$^2$*)* and $h$ (*m*) are the acceleration due to gravity and the lava flow thickness, respectively. The acceleration due to gravity on Mercury is $g$=3.61 m/s$^2$ (i.e. almost 1/3 of that of the Earth; *Mazarico et al*. 2014). $\lambda$ can be calculated as follows:

$$\lambda = \frac{1}{[0.79\ln(Re) - 1.64]^2} \quad (3)$$

where *Re* is the Reynolds numbers defined as:

$$Re = \frac{2\rho uh}{\eta} \quad (4)$$

ρ is the bulk lava density (2450 kg/m$^3$ for the melt considered here, calculated following *Ochs and Lange* [1999]) and $\eta$ is the lava viscosity (Pa s).

With regard to the thickness of the flows to be used in the above equations, the total thickness of NVP lavas has been estimated to be ~ 0.7–1.8 km [*Ostrach et al*., 2015; *Head et al*., 2011; *Klimczak et al*., 2012; *Byrne et al*., 2013]. However, this thickness is likely to be the result of the superimposition of different lava flows, whose individual thicknesses are presently unknown. Some constraints can be derived from terrestrial analogous lavas. Among them, Hawaiian lavas with rheological behaviours similar to the silicate melt considered here typically show individual flow thicknesses ranging from 1 to 5 m [*Griffiths*, 2000]; furthermore, komatiites, also commonly considered similar to Mercurian lavas [*Weider et al*., 2012], show flows with a slightly greater thickness of about 10 m [*Williams et al*., 2001]. Accordingly, values of $h$ from 1 to 10 m are used in equations 2 and 4.

The values of viscosity to be used in the above equations depend upon the eruptive temperature. Possible eruption temperatures for Mercurian lavas are estimated to be around 1623 K [*Charlier et al*., 2013; *Namur et al*., 2016b]. As the estimated liquidus temperature for the melt used in the experiments is similar (1600 K) to the suggested eruptive temperature, we started our modelling considering 1623 K as a representative eruptive temperature of the lava that is initially erupted. Changing the eruptive temperature to 1600 K does not affects the outcomes of the model presented below. According to our experimental data and modelling (Figure 2), at 1623 K the viscosity of the silicate melt is ca. 13 Pa s. Assuming constant effusion rates, calculated velocities for lava thicknesses of 1, 5 and 10 m, considering varying ground slopes (from 0.1 to 5°) are shown in **Figure 11**. As expected, results show that lava flows with larger thickness consistently have larger velocities (see also Table S3 in the supporting materials). In addition, in order to maintain the turbulent regime (*Re* larger than 2000) for lava emplacement (*Byrne et al*., 2013), minimum velocities must be larger than ca. 1.2 m/s (on a slope of 0.3°), and ca. 0.7 m/s (on a slope of 0.1°) for lava thicknesses of 5 and 10 m, respectively. Note that for lava thicknesses on the order of 1.0 m, the turbulent regime is only possible for slopes larger than 10°, a value which appears unlikely on the basis of recent studies in this sector of the planet's surface [e.g., *Byrne et al*., 2013; *Hurwitz et al*., 2013]. As the slope increases, the lava flow velocity can increase up to values of the order of 6.5 and 7.2 m/s, when flowing on a 5° ground slope, for thicknesses of 5 and 10 m, respectively.

In order to incorporate the possible effects of heat loss during the lava flow in our model, we performed thermal balance calculations using the FLOWGO model [*Harris and Rowland*, 2001]. In the model we consider that lavas can flow in channels with variable widths from 100, 1000 and 30000 m, as observed for the NVP on Mercury [*Byrne et al*., 2013]. In addition, we consider, as minimum and maximum velocities of the lava, the values obtained above for a lava flow of thickness $h$=10 m flowing on a slope of 0.1° ($u$=0.7 m/s) and for a lava flow of the same thickness flowing on a slope of 5° ($u$=7.2 m/s). Note that, according to the above discussion, these end-member values of $u$ include all possible values of velocity for a lava flowing down a slope at an angle between 0.1 and 5°, with a thicknesses ranging from 5–10 m. A further constraint to be considered in the model is the emissivity value of the lava. According to *Harris* [2013] this parameter shows little variability, even among very different magmatic compositions. In particular, it ranges from ca. 0.8 to ca. 0.9 from basalt to trachyte; accordingly, in the model, we used an emissivity value of 0.85. The differences produced when this parameter is changed are negligible. The FLOWGO model accounts for the formation of a crust developing on the outer part of the lava acting as a thermal insulation boundary and limiting the heat loss. In addition, turbulence cannot be directly included in FLOWGO. However, the large velocity values used in the model can be considered as a proxy for turbulence implying that larger velocities produce larger covered areas by the lava and, hence, larger heat losses. Parameters used in the model are reported in Table S4 (supplementary materials); details about the FLOWGO model can be found in *Harris and Rowland* [2001].

**Figure 12** shows the variation of heat loss (in K/km) of the lava as a function of effusion rate (in $m^3$/s). The plot displays six curves corresponding to lava flows with the above-considered velocities (0.7 and 7.2 m/s) flowing in channels with widths of 100, 1000 and 30000 m. The graph shows that the curves have a similar behaviour and tend to saturate towards constant values of heat loss at high effusion rates (of the order of $10^4$, $10^5$ and $10^7$ for channels having width of 100, 1000 and 30000 m, respectively). At lower effusion rates the heat loss dramatically increases (Figure 12). The curves corresponding to the lava flowing with a velocity of 0.7 m/s indicate that, at the same value of effusion rate, the heat loss is always lower in comparison to the lava flowing with a velocity of 7.2 m/s. From Figure 12, it also emerges that the channel width plays a major role in modulating heat loss. In fact, to maintain a similar amount of heat loss, strongly increasing effusion rates are required as the width of the channel increases.

As stressed above, a fundamental constraint to be satisfied is that a turbulent regime likely characterized the emplacements of the lava for distances on the order of 100 km [*Byrne et al.*, 2013]. According to the above discussion, Reynolds numbers larger than 2000 are possible considering the eruptive temperature T=1623 K (and relative value of viscosity, $\eta$=13 Pa s), and lava velocities larger than 0.7 m/s, according to the slope values and lava thicknesses reported above (see supporting information, Table S3). Keeping lava velocity constant in the limits imposed above (i.e. 0.7 and 7.2 m/s), we can use the results shown in Figure 12 to evaluate how heat loss impacts lava rheology and the consequent dynamic regime of lava emplacement. As the lava cools down during its flow, its viscosity is expected to increase leading to a decrease of the Reynolds number and resulting, eventually, in the suppression of the turbulent regime. According to our experimental results and modelling (Figure 2), the viscosity of the analogous melt studied here shows minimal variations (13–20 Pa s) in the temperature range 1623–1581 K (i.e. from eruptive to liquidus temperature; Figure 3). This implies that the Reynolds number remains relatively constant during this temperature drop ($\Delta$T=42 K, i.e. 1623–1581 K) allowing the turbulent regime to remain almost unchanged. Considering that the lava has to cover a distance on the order of 100 km in turbulent regime [*Byrne et al.*, 2013], this would correspond to a heat loss of ca. 0.4 K/km. According to the model shown in Figure 12, a heat loss of 0.4 K/km can be obtained at different effusion rates depending on the width of the channel in which the lava flows. In particular, effusion rates larger than ca. $10^4$, $10^5$ and $10^6$ $m^3/s$ are required for channel widths of 100, 1000 and 30000m, respectively. It is noteworthy that these estimates are in concurrence with independent results given by *Keszthelyi and Self* [1998] for the emplacement of long (on the order of 100 km) basaltic lava flows. In particular, *Keszthelyi and Self* [1998] report that in order to cover these large distances, a heat loss equal to or lower than 0.5 K/km is necessary. Furthermore, these authors suggest that effusion rates larger than $10^3$–$10^4$ $m^3/s$, and velocities on the order of 4–12 m/s are also required. These values are comparable to those arising from our model. From this discussion, it is therefore clear that the Mercurian analog melt studied here is able to flow over the long distances observed for the NVP in turbulent regime, under the effusion rates given above. Significantly, the required effusion rates (and relative single lava flow thickness) are comparable to those estimated for some of Earth's basaltic eruptions forming the so-called LIPs [e.g. *Bryan et al.*, 2010; *Head et al.* 2011]. Consequently, the mechanism of eruption and emplacement of terrestrial flood basalts can be considered, as a first approximation, as a plausible geologic analogue for the Mercury's NVP lavas in agreement with the results reported by Vander Kaaden and McCubbin [2016].

A further issue that might deserve consideration is the possible effect of cooling of the lavas due to the temperature difference between diurnal and nocturnal times on Mercury. During a Mercurian day (corresponding to 59 terrestrial days), surface temperature can reach 725 K. At night, the surface temperature drops to about 90 K [*Strom,* 1997]. In these conditions, the high temperature of the planet may limit heat dissipation during daytime eruptions and, consequently, enhance the emplacement of lava flows. Conversely, during the night, the temperature drop could act as a limiting factor for the distance the lava is able to flow. However, in modelling lava flows, the formation of a crust developing on the outer part of the lava must be considered. This effect is considered in the FLOWGO models presented above. The formation of this crust acts as thermal insulation for the lava and, therefore, the effect of temperature difference between Mercurian day and night can be considered negligible.

According to the above discussion, it is apparent that NVP lavas were able to cover large distances without undergoing strong degrees of crystallization. Accordingly, most of their solidification history would have occurred when they stopped flowing and emplaced, defining the present morphology of Mercury's NVP. The causes why lavas stopped flowing on the surface of the planet can be variable. One possibility might be that the topography of the planet played a key role, for example because of the presence of low topographies due to pre-existing impact craters that allowed the lavas to stagnate, lose heat and solidify.

Further evidence that solidification of the lavas must have occurred mostly after emplacement is provided by results from crystallization experiments. Lavas can strongly reduce their ability to flow when approaching the so-called "maximum packing fraction". The maximum packing fraction strongly depends on the aspect ratio of crystals [*Mader et al.,* 2013]. Our experiments indicate that crystal aspect ratios range, on average, between 2 and 14. According to *Mader et al.* [2013] this corresponds to maximum packing fractions from ca. 52% to 28%. As here we are evaluating the least favourable conditions under which the lavas can flow, we considered the lowest maximum packing fraction (i.e. 28%) as the reference crystal fraction that would strongly reduce (or stop) the ability of the lava to flow. Our experiments (Figure 4a-b) indicate that temperature needs to drop to 1502 K to reach a crystal content of ca. 28%. The question is whether the lava in these conditions can preserve the turbulent flow regime, as required by morphological constraints provided by [*Byrne et al.*, 2013]. Our experimental results presented in Figure 4 and Table 3 show that the rate of increase in viscosity due to crystallization is different depending on the applied shear rate, due to a shear thinning behaviour of the lava. As an example, at T=1520 K (corresponding to a crystal content of ca. 10%; Table 3) viscosity changes from 144 Pa s to 1259 Pa s, as the shear rate decreases from 5 to 0.1 $s^{-1}$. A similar behavior is observed for lower temperatures, i.e. larger crystal contents, up to the threshold limit of 28% crystal content (Figure 4).

The Reynolds number (equation 4) can be used to assess whether the lava can emplace in turbulent flow conditions, i.e. *Re* larger than 2000. In the following calculations, we consider a lava flow with thickness of 10 m flowing at velocities from 0.7 to 7.2 m/s (see above) and having viscosities determined by the amount of crystals formed upon solidification (Figure 4). Note that for lavas with thicknesses lower than ca. 10 m, considering the viscosities for crystal-bearing lavas with a crystallinity equal to or larger than 5% (Figure 4 and Table 3) and for any of the above velocities, *Re* is always lower than 2000; in these conditions the laminar regime prevails. For lava thicknesses on the order of 10 m, i.e. maximum thickness considered here, the laminar fluid dynamic regime governs most of the behavior of the flowing lava. However, according to our experimental results, there are cases in which the turbulent regime might still persist, even for partially crystallized systems. In particular, *Re* can attain values above 2000 if viscosity is lower than 144 Pa s and flow velocity is larger than 6 m/s. From Figure 4, these viscosity values correspond to partially crystallized systems with crystal contents up to ca. 10% for shear rates of 5 $s^{-1}$. Lower shear rates shift *Re* towards values lower than 2000, driving the system towards laminar conditions. These considerations highlight that, whilst for some narrow combinations of parameters (i.e. viscosity, velocity, lava thickness), the lava could potentially flow in turbulent conditions below the liquidus temperature ($T_L$), in most cases it behaves as a laminar system. However, this is contrary to morphological features observed from satellite images and considered to reflect a turbulent emplacement of the lava [*Byrne et al.*, 2013]. These considerations corroborate the idea proposed above, that crystallization of lavas is a process that most likely occurred after emplacement.

## 6. Conclusions

The new experiments and modelling presented in this work allow us to shed new light on the mechanisms that determined the emplacement of what is believed to be one the largest volcanic deposits in the Solar System [*Byrne et al.*, 2016]; the northern volcanic province on planet Mercury. The high $Na_2O$ content (~8.8 wt.%) of the experimental starting material plays an important role in reducing lava viscosity, as confirmed by concentric cylinder high temperature viscosity measurements.

The viscosity of our Mercury analog silicate melt measured at superliquidus temperature conditions slightly increases from 4 to 16 Pa s in the temperature range 1736–1600 K. In the temperature range 1569–1502 K (subliquidus), viscosity increases due to the combined effect of progressive crystallization (from 2 to 28 area %) and chemical evolution of the melt. Here, a shear-thinning behavior was observed when varying strain rates from 0.1 and 5 $s^{-1}$. Lava viscosity decreases by ca. 1 log unit as shear rate varies from 0.1 and 5 $s^{-1}$.

These new viscosity measurements were used to model the behaviour of Mercurian lavas during emplacement. Merging experimental data and numerical modelling leads to the conclusion that the emplacement of lavas in turbulent conditions, as claimed by previous works [e.g. *Byrne et al.*, 2013], defines a geologic scenario in which lavas might have travelled long distances without undergoing strong degrees of crystallization. Therefore, it is possible to infer that solidification (crystallization) of the lavas mostly occurred after emplacement on the surface of the planet. Effusion rates were estimated to be in the order of $10^4$–$10^7$ $m^3/s$, comparable to those estimated for some Earth's basaltic eruptions forming the LIPs.

## Acknowledgments

This work was funded by the European Research Council for the Consolidator Grant ERC-2013-CoG Proposal No. 612776 — CHRONOS to Diego Perugini. Francesco Vetere acknowledges support from the TESLA FRB-base project from the Department of Physics and Geology, University of Perugia. Olivier Namur acknowledges support from an Individual Intra-European Marie Curie Fellowship (SULFURONMERCURY) and from the DFG through the Emmy Noether Program. Rebecca Astbury is gratefully acknowledged for language corrections. We also wish to thank Nonna Rita for her wise words "adduvi c'è gustu, un c'è pirdenza" that guided us through the writing of this work. Data of experiments are reported in Tables 1, 2 and 3, in Figures 1–12 and in Supporting Information as Tables S1, S2, S3 and S4 and Figure S1.

Supporting Information for

# Experimental constraints on the rheology, eruption and emplacement dynamics of analog lavas comparable to Mercury's northern volcanic plains

**F. Vetere[1], S. Rossi[1], O. Namur[2,3], D. Morgavi[1], V. Misiti[4], P. Mancinelli[1], M. Petrelli[1], C. Pauselli[1], D. Perugini[1]**

[1] Department of Physics and Geology, University of Perugia, Perugia, Italy.

[2] Institut für Mineralogie, Leibniz Universität Hannover, Hannover, Germany.

[3] Department of Earth and Environmental Sciences, KU Leuven, Leuven, Belgium

[4] Istituto Nazionale di Geofisica e Vulcanologia, Roma, Italy.

## *Introduction*

Here we report detailed description on data presented in the main text. In particular, a description of the theoretical and experimental approaches on two phase magma rheology are reported in Text S1. Moreover, details on the a) chemistry for the residual glasses after crystallization processes, b) the chemistry of the solid phases produced and for c) the calculation concerning the velocities and relative Reynolds number used in this study are provided in an Excel file (Tables S1, S2 and S3, respectively)

Text S1.

In order to quantify the effect of cooling and relative crystals content, size distributions, and crystal shapes on magma viscosity, abundant theoretical and experimental studies were reported [Shaw, 1969; Murase and McBirney, 1973; Murase et al., 1985; Ryerson et al., 1988; Pinkerton and Stevenson, 1992; Lejeune and Richet, 1995; Pinkerton and Norton, 1995; Sato, 2005; Arbaret et al., 2007; Caricchi et al., 2007; Ishibashi and Sato, 2007; Costa et al., 2009; Vetere et al., 2010, 2013, Vona et al., 2011, 2013; Sehlke et al., 2014; Sehlke and Whittington, 2015]. As described by [Ishibashi, 2009], one of the most frequently used equations to evaluate the relative viscosity of a suspension $\eta r$ ($\eta r = \eta eff/\eta m$), where $\eta eff$ is the effective viscosity of the suspension with a volume fraction of crystals, and ηm is the viscosity of the melt, is the Krieger–Dougherty (KD) equation (Krieger and Dougherty, 1959):

$$\eta_r = (1-\Phi/\ \Phi_m)^{-\nu\Phi m} \quad (1)$$

where Φ is a volume fraction of suspended particles, Φm is the maximum packing density, and ν is the so called intrinsic viscosity (a measure of the crystal influence to the magmatic system viscosity). Assuming that νΦm = 2.5, the effective viscosity of crystal + melts systems is often estimated with the Einstein-Roscoe equation (Einstein, 1906; Roscoe, 1952):

$$\eta_{eff} = \eta_m\ (1-\Phi/\ \Phi_m)^{-2.5} \quad (2)$$

However, this equation is only valid if the shape of the particles can be approximated to that of a sphere. The Φm value, according to Marsh [1981], is estimated to be 0.6 and corresponds to the crystal fraction at which a transition of the system to the rigid solid state occurs. However, it has been demonstrated that this value is not always adequate for natural magmatic systems [Costa, 2005].

Studies on the viscosity of magmatic suspensions can be done allowing crystals to nucleate and grow from melts at subliquidus conditions in order to examine the magma's rheological evolution [Pinkerton and Norton, 1995; Sato, 2005; Ishibashi and Sato, 2007; Ishibashi, 2009, Vona et al., 2011, Vetere et al., 2013]. This approach has the advantage of tracking the sequential variation of both the crystal texture and rheological properties during cooling of magmas and is adopted here.

A fluid is said to be purely viscous if the shear stress (σ) is a function only of the shear rate ($\dot{\gamma}$). For non-Newtonian fluids in simple shear flow a viscosity function $\eta(\dot{\gamma})$ is:

$$\eta(\dot{\gamma}) = \tau/\dot{\gamma} \quad (3)$$

The viscosity function is also called the apparent viscosity (this term is used to indicate that the result depends on the particular strain rate at which it was measured). Rheological parameters of two-phase suspension are usually treated by using classical approach for pseudoplastic (power law) flow.

By using such approach it is possible to identify parameters such as the flow index n and consistency K by the following:

$$\sigma = K\ \dot{\gamma}^{\,n} \quad (4)$$

For a Newtonian liquid, n =1. More details can be found in Moitra and Gonnermann (2014) for the Herschel-Bulkley model (Herschel and Bulkley, 1926). Then, fitting can be expressed in ln or log terms for stress or for apparent viscosity:

$$\ln \sigma = \ln k + n \ln \dot{\gamma} \quad (5)$$

and/or

$$\log \eta = \log k + (n-1) \log \dot{\gamma} \quad (6)$$

Tables

Table 1

| | This work | | SW-2015 | | | |
|---|---|---|---|---|---|---|
| | wt% | std | Enstatite bas | NVP | NVP-Na | IcP-HCT |
| $SiO_2$ | 61.48 | 0.36 | 55.06 | 57.10 | 55.02 | 53.30 |
| $TiO_2$ | 0.36 | 0.01 | 0.18 | 0.96 | 0.89 | 0.89 |
| $Al_2O_3$ | 8.95 | 0.11 | 13.07 | 15.27 | 14.88 | 12.31 |
| FeO | - | - | 0.29 | 3.61 | 2.88 | 3.31 |
| MnO | - | - | 0.14 | 0.25 | 0.25 | 0.23 |
| MgO | 14.12 | 0.21 | 19.72 | 16.53 | 13.59 | 22.53 |
| CaO | 6.81 | 0.09 | 12.37 | 4.95 | 4.29 | 6.29 |
| $Na_2O$ | 8.85 | 0.21 | 0.04 | 0.29 | 6.25 | 0.16 |
| $K_2O$ | 0.21 | 0.02 | 0.08 | 0.31 | 0.22 | 0.19 |
| | | | | | | |
| Tot | 100.78 | | 100.95 | 99.27 | 98.27 | 99.21 |
| NBO/T | 0.89 | | 1.00[a] | 0.64[a] | 0.68[a] | 1.05[a] |
| | | | – | 0.50[b] | 0.55[b] | 0.86[b] |
| | | | – | 0.41[c] | 0.53[c] | 0.79[c] |

**Table 1.** Electron microprobe analyses of the starting material compared to literature data. The starting composition represents an average of 50 measurements on the synthetic starting glass material. SW-2015 refers to literature data presented in *Sehlke and Whittington* [2015]. Std refers to the standard deviations. The NBO/T values (non-bridging oxygens (NBO) per tetrahedrally coordinated cation (T) [*Mysen and Richet*, 2005]). NBO/T values for *Sehlke and Whittington* [2015] chemical compositions refer to calculation results considering (a) ferrous only, (b) ferrous and ferric, and (c) ferric only.

Table 2

| # | T (K) | η (Pa s) | std | $\dot{\gamma}$ |
|---|---|---|---|---|
| M29 | 1600 | 16.3 | 0.27 | 5 |
| M28 | 1610 | 14.9 | 0.58 | 5 |
| M32 | 1618 | 12.6 | 0.43 | 5 |
| M27 | 1621 | 14.1 | 0.41 | 5 |
| M26 | 1630 | 12.9 | 0.35 | 5 |
| M31 | 1639 | 10.3 | 0.06 | 5 |
| M15 | 1652 | 9.7 | 0.05 | 5 |
| M14 | 1662 | 8.7 | 0.07 | 5 |
| M13 | 1671 | 7.9 | 0.04 | 5 |
| M12 | 1681 | 7.2 | 0.05 | 5 |
| M17 | 1687 | 6.7 | 0.11 | 5 |
| M10a | 1691 | 6.8 | 0.05 | 5 |
| M10b | 1691 | 6.7 | 0.08 | 5 |
| M9 | 1701 | 6.0 | 0.06 | 5 |
| M7a | 1711 | 5.5 | 0.05 | 5 |
| M7b | 1711 | 5.7 | 0.03 | 5 |
| M 6 | 1720 | 5.0 | 0.03 | 5 |
| M 5 | 1730 | 4.6 | 0.04 | 5 |
| M 3a | 1736 | 4.3 | 0.03 | 5 |
| M 3b | 1736 | 4.0 | 0.04 | 5 |
| M 16a | 1642 | 10.7 | 0.17 | 5 |
| M 16b | 1642 | 11.2 | 0.15 | 5 |
| M 16c | 1642 | 11.2 | 0.26 | 6 |
| M 16d | 1642 | 11.3 | 0.23 | 7 |
| M16e | 1642 | 11.2 | 0.25 | 8 |
| M16f | 1642 | 11.4 | 0.21 | 9 |
| M16g | 1642 | 11.2 | 0.16 | 10 |

**Table 2.** Experimental conditions and results of viscosity measurements (η) using the concentric cylinder apparatus. $\dot{\gamma}$ (in $s^{-1}$) refers to the applied shear stress; std refers to standard deviation of viscosity measurements.

Table 3

| # | Temperature | Log η Pa s | Log η Pa s | Log η Pa s | Flow index | $K$ | Olivine | Pyroxene | Crystallinity | Crystallinity mass balance |
|---|---|---|---|---|---|---|---|---|---|---|
| | K | $\dot{\gamma}$=5 $s^{-1}$ | $\dot{\gamma}$=1 $s^{-1}$ | $\dot{\gamma}$=0.1 $s^{-1}$ | $n$ | Pa s | area % | area % | area % | vol % |
| M33 | 1502 | 3.04 | 3.36 | 4.01 | 0.4201 | 1172± 175 | 17.6 ± 3.0 | 9.9 ± 1.3 | 27.5 ± 3.0 | 23.7 |
| M34 | 1510 | 2.67 | 3.18 | 3.59 | 0.4551 | 593± 53 | 11.5 ± 1.9 | 9.9 ± 0.8 | 21.4 ± 2.2 | 16.3 |
| M35 | 1520 | 2.16 | 2.69 | 3.10 | 0.4568 | 207± 30 | 9.9 ± 0.2 | 1.7 ± 0.2 | 11.8 ± 1.2 | 10.1 |
| M36 | 1533 | 1.99 | 2.47 | 2.72 | 0.5272 | 131± 22 | 8.8 ± 0.9 | - | 8.8 ± 0.9 | 8.2 |
| M37 | 1545 | 1.93 | 2.25 | | 0.5333 | 61± 5 | 5.0 ± 0.7 | - | 5.0 ± 0.7 | 7.3 |
| M38 | 1569 | 1.63 | | | - | - | 2.1 ± 0.3 | - | 2.1 ± 0.3 | 3.9 |

**Table 3:** Experimental conditions and results of viscosity measurements (η) during crystallization experiments at different shear rate ($\dot{\gamma}$). Rheological parameters [flow index ($n$) and consistency ($k$)] and crystal contents (both area% and vol.%) are also reported.

## Table S1

| $\gamma = 0.1\ s^{-1}$ | 1545 K | Std | 1533 K | Std | 1520 K | Std |
|---|---|---|---|---|---|---|
| SiO2 | 61.94 | 0.44 | 63.03 | 0.60 | 62.95 | 0.43 |
| TiO2 | 0.40 | 0.02 | 0.42 | 0.03 | 0.43 | 0.02 |
| Al2O3 | 9.74 | 0.16 | 9.78 | 0.12 | 10.16 | 0.16 |
| MgO | 10.59 | 0.20 | 9.98 | 0.21 | 9.16 | 0.16 |
| CaO | 7.30 | 0.09 | 7.42 | 0.14 | 7.02 | 0.09 |
| Na2O | 8.46 | 0.15 | 8.66 | 0.19 | 8.76 | 0.20 |
| K2O | 0.21 | 0.02 | 0.22 | 0.01 | 0.21 | 0.02 |
| | | | | | 98.70 | |

| $\gamma = 1\ s^{-1}$ | 1545 K | Std | 1533 K | Std | 1520 K | Std | 1510 K | Std | 1502 K | Std |
|---|---|---|---|---|---|---|---|---|---|---|
| SiO2 | 62.85 | 0.54 | 63.30 | 0.32 | 64.76 | 0.57 | 65.05 | 0.58 | 65.95 | 0.37 |
| TiO2 | 0.39 | 0.02 | 0.40 | 0.02 | 0.43 | 0.02 | 0.44 | 0.02 | 0.47 | 0.02 |
| Al2O3 | 9.56 | 0.09 | 9.63 | 0.09 | 10.56 | 0.11 | 10.81 | 0.13 | 11.58 | 0.16 |
| MgO | 10.54 | 0.15 | 10.12 | 0.19 | 8.69 | 0.18 | 8.14 | 0.20 | 7.18 | 0.17 |
| CaO | 7.29 | 0.12 | 7.34 | 0.11 | 6.56 | 0.08 | 6.15 | 0.08 | 4.93 | 0.08 |
| Na2O | 9.54 | 0.19 | 9.42 | 0.26 | 8.75 | 0.40 | 8.33 | 0.43 | 9.00 | 0.28 |
| K2O | 0.22 | 0.01 | 0.22 | 0.01 | 0.25 | 0.01 | 0.25 | 0.02 | 0.27 | 0.02 |

| $\gamma = 5\ s^{-1}$ | 1569 K | Std | 1545 K | Std | 1533 K | Std | 1520 K | Std | 1510 K | Std | 1502 K | Std |
|---|---|---|---|---|---|---|---|---|---|---|---|---|
| SiO2 | 61.68 | 0.26 | 62.67 | 0.24 | 63.42 | 0.27 | 63.24 | 0.27 | 64.30 | 0.34 | 64.64 | 0.40 |
| TiO2 | 0.40 | 0.04 | 0.40 | 0.04 | 0.44 | 0.04 | 0.43 | 0.02 | 0.46 | 0.04 | 0.49 | 0.04 |
| Al2O3 | 8.97 | 0.12 | 9.73 | 0.16 | 9.43 | 0.13 | 9.58 | 0.15 | 10.27 | 0.12 | 11.11 | 0.11 |
| MgO | 11.89 | 0.23 | 10.35 | 0.32 | 10.03 | 0.19 | 9.33 | 0.10 | 8.69 | 0.23 | 7.51 | 0.19 |
| CaO | 6.84 | 0.10 | 7.12 | 0.11 | 7.20 | 0.12 | 7.08 | 0.09 | 6.21 | 0.11 | 4.94 | 0.09 |
| Na2O | 7.99 | 0.13 | 8.11 | 0.13 | 8.29 | 0.15 | 8.38 | 0.14 | 9.15 | 0.13 | 9.69 | 0.09 |
| K2O | 0.19 | 0.03 | 0.20 | 0.02 | 0.24 | 0.03 | 0.20 | 0.02 | 0.25 | 0.02 | 0.25 | 0.03 |
| P2O5 | 0.00 | 0.00 | 0.00 | 0.00 | 0.00 | 0.00 | 0.00 | 0.00 | 0.00 | 0.00 | 0.00 | 0.00 |

**$\gamma$ = shear rate**

**Table S1.** Electron microprobe analyses of the residual glasses material provided in Figure 6 after crystallization experiments. Presented data are averages of 50 measurements on each experimental sample. Std refers to the standard deviations.

## Table S2

| $\gamma = 0.1\ s^{-1}$ | Ol 1545 K | Std | Ol 1533 K | Std | CPx 1520 K | Std | Ol 1520 K | Std |
|---|---|---|---|---|---|---|---|---|
| **SiO2** | 42.54 | 0.63 | 42.15 | 0.85 | 55.75 | 0.78 | 42.42 | 0.73 |
| **TiO2** | 0.02 | 0.01 | 0.01 | 0.01 | 0.09 | 0.01 | 0.01 | 0.01 |
| **Al2O3** | 0.02 | 0.03 | 0.01 | 0.01 | 0.42 | 0.03 | 0.02 | 0.01 |
| **MgO** | 56.40 | 1.15 | 57.11 | 0.71 | 22.26 | 0.65 | 57.19 | 1.56 |
| **CaO** | 0.32 | 0.03 | 0.27 | 0.07 | 20.82 | 0.80 | 0.30 | 0.04 |
| **Na2O** | 0.04 | 0.02 | 0.01 | 0.01 | 0.30 | 0.04 | 0.04 | 0.03 |
| **K2O** | 0.01 | 0.01 | 0.01 | 0.01 | 0.02 | 0.02 | 0.02 | 0.01 |

| $\gamma = 1\ s^{-1}$ | Ol 1545 K | Std | Ol 1533 K | Std | CPx 1520 K | Std | Ol 1520 K | Std | CPx 1510 K | Std | Ol 1510 K | Std | CPx 1502 K | Std | Ol 1502 K | Std |
|---|---|---|---|---|---|---|---|---|---|---|---|---|---|---|---|---|
| **SiO2** | 42.69 | 0.39 | 42.93 | 0.62 | 56.37 | 0.18 | 42.71 | 0.40 | 56.27 | 0.47 | 43.17 | 0.53 | 56.42 | 0.34 | 43.21 | 0.39 |
| **TiO2** | 0.00 | 0.01 | 0.01 | 0.01 | 0.10 | 0.02 | 0.00 | 0.01 | 0.11 | 0.03 | 0.00 | 0.01 | 0.10 | 0.01 | 0.00 | 0.01 |
| **Al2O3** | 0.01 | 0.02 | 0.03 | 0.04 | 0.45 | 0.04 | 0.02 | 0.04 | 0.48 | 0.07 | 0.05 | 0.11 | 0.49 | 0.11 | 0.02 | 0.02 |
| **MgO** | 57.08 | 0.39 | 57.09 | 0.52 | 22.68 | 0.71 | 57.09 | 0.57 | 22.08 | 0.63 | 56.93 | 0.52 | 23.10 | 0.82 | 57.38 | 0.26 |
| **CaO** | 0.30 | 0.03 | 0.32 | 0.04 | 20.21 | 0.80 | 0.32 | 0.04 | 20.85 | 0.49 | 0.34 | 0.07 | 19.62 | 0.94 | 0.31 | 0.03 |
| **Na2O** | 0.04 | 0.03 | 0.07 | 0.02 | 0.28 | 0.04 | 0.08 | 0.05 | 0.28 | 0.06 | 0.08 | 0.08 | 0.34 | 0.09 | 0.08 | 0.02 |
| **K2O** | 0.02 | 0.01 | 0.01 | 0.01 | 0.02 | 0.01 | 0.02 | 0.01 | 0.01 | 0.01 | 0.02 | 0.01 | 0.02 | 0.01 | 0.02 | 0.01 |

| $\gamma = 5\ s^{-1}$ | Ol 1569 K | Std | Ol 1545 K | Std | Ol 1533 K | Std | Cpx 1520 K | Std | Ol 1520 K | Std | Cpx 1510 K | Std | Ol 1510 K | Std | Cpx 1502 K | Std | Ol 1502 K | Std |
|---|---|---|---|---|---|---|---|---|---|---|---|---|---|---|---|---|---|---|
| **SiO2** | 42.26 | 0.29 | 41.73 | 0.36 | 42.32 | 0.34 | 56.22 | 1.40 | 42.18 | 0.17 | 56.63 | 0.63 | 41.23 | 0.65 | 56.78 | 0.60 | 41.81 | 0.12 |
| **TiO2** | 0.00 | 0.00 | 0.04 | 0.05 | 0.04 | 0.03 | 0.07 | 0.00 | 0.02 | 0.01 | 0.12 | 0.04 | 0.00 | 0.00 | 0.13 | 0.05 | 0.01 | 0.00 |
| **Al2O3** | 0.03 | 0.02 | 0.04 | 0.03 | 0.03 | 0.03 | 0.46 | 0.06 | 0.10 | 0.09 | 1.20 | 0.61 | 0.02 | 0.03 | 1.15 | 0.65 | 0.09 | 0.04 |
| **MgO** | 54.39 | 0.80 | 56.16 | 0.57 | 54.75 | 0.96 | 22.92 | 0.95 | 55.24 | 0.61 | 21.65 | 1.52 | 54.03 | 1.15 | 21.94 | 1.58 | 54.87 | 0.52 |
| **CaO** | 0.27 | 0.05 | 0.31 | 0.03 | 0.32 | 0.04 | 19.65 | 0.03 | 0.35 | 0.66 | 18.75 | 0.61 | 0.36 | 0.06 | 18.47 | 0.90 | 0.35 | 0.04 |
| **Na2O** | 0.02 | 0.01 | 0.02 | 0.01 | 0.06 | 0.04 | 0.25 | 0.07 | 0.05 | 0.03 | 0.92 | 0.53 | 0.11 | 0.04 | 0.80 | 0.43 | 0.08 | 0.03 |
| **K2O** | 0.01 | 0.01 | 0.02 | 0.01 | 0.03 | 0.04 | 0.02 | 0.01 | 0.01 | 0.02 | 0.06 | 0.03 | 0.06 | 0.01 | 0.02 | 0.01 | 0.02 | 0.00 |

**γ = shear rate**

**Table S2** Electron microprobe analyses of the crystalline phases, for experiments M33– M37 (Table 3) in temperature range 1569–1502 K. Std refers to the standard deviations.

**Table S3**
**Data derived by using equation 2, 3 and 4. Please refer to text**

| | | | | | | | |
|---|---|---|---|---|---|---|---|
| gravity | $m/s^2$ | | 3.6 | | 3.6 | | 3.6 | |
| lava flow thickness | (m) | | 10 | | 5 | | 1 | |
| lava bulk density | $(Kg/m^3)$ | | 2452 | | 2452 | | 2452 | |
| dynamic viscosity | Pa s | | 13 | | 13 | | 13 | |

| slope (°) | velocity (m/s) | Re | velocity (m/s) | Re | velocity (m/s) | Re |
|---|---|---|---|---|---|---|
| 0.1 | 0.732 | 2762 | 0.611 | 1153 | 0.353 | 133 |
| 0.2 | 1.109 | 4183 | 0.973 | 1836 | 0.590 | 223 |
| 0.3 | 1.410 | 5320 | 1.227 | 2315 | 0.786 | 296 |
| 0.4 | 1.671 | 6304 | 1.459 | 2751 | 0.955 | 360 |
| 0.5 | 1.905 | 7186 | 1.677 | 3164 | 1.111 | 419 |
| 0.6 | 2.119 | 7995 | 1.873 | 3532 | 1.254 | 473 |
| 0.7 | 2.319 | 8748 | 2.048 | 3862 | 1.388 | 524 |
| 0.8 | 2.507 | 9456 | 2.214 | 4177 | 1.515 | 572 |
| 0.9 | 2.685 | 10127 | 2.377 | 4484 | 1.636 | 617 |
| 1.0 | 2.854 | 10766 | 2.534 | 4780 | 1.751 | 661 |
| 1.1 | 3.016 | 11378 | 2.682 | 5058 | 1.862 | 703 |
| 1.2 | 3.172 | 11967 | 2.821 | 5320 | 1.970 | 743 |
| 1.3 | 3.323 | 12534 | 2.955 | 5573 | 2.073 | 782 |
| 1.4 | 3.468 | 13083 | 3.087 | 5823 | 2.174 | 820 |
| 1.5 | 3.609 | 13615 | 3.217 | 6069 | 2.271 | 857 |
| 1.6 | 3.746 | 14132 | 3.344 | 6307 | 2.366 | 893 |
| 1.7 | 3.879 | 14634 | 3.465 | 6535 | 2.459 | 928 |
| 1.8 | 4.009 | 15125 | 3.581 | 6754 | 2.549 | 962 |
| 1.9 | 4.136 | 15603 | 3.695 | 6969 | 2.638 | 995 |
| 2.0 | 4.260 | 16071 | 3.808 | 7182 | 2.724 | 1028 |
| 2.1 | 4.381 | 16528 | 3.920 | 7394 | 2.809 | 1060 |
| 2.2 | 4.500 | 16976 | 4.030 | 7600 | 2.893 | 1091 |
| 2.3 | 4.617 | 17415 | 4.136 | 7801 | 2.974 | 1122 |
| 2.4 | 4.731 | 17846 | 4.238 | 7994 | 3.055 | 1152 |
| 2.5 | 4.843 | 18269 | 4.339 | 8185 | 3.133 | 1182 |
| 2.6 | 4.953 | 18685 | 4.440 | 8374 | 3.211 | 1211 |
| 2.7 | 5.062 | 19094 | 4.540 | 8562 | 3.288 | 1240 |
| 2.8 | 5.168 | 19496 | 4.638 | 8749 | 3.363 | 1269 |
| 2.9 | 5.273 | 19892 | 4.734 | 8930 | 3.437 | 1297 |
| 3.0 | 5.376 | 20282 | 4.828 | 9106 | 3.510 | 1324 |
| 3.1 | 5.478 | 20666 | 4.919 | 9279 | 3.582 | 1351 |
| 3.2 | 5.579 | 21045 | 5.011 | 9451 | 3.654 | 1378 |
| 3.3 | 5.678 | 21419 | 5.102 | 9624 | 3.724 | 1405 |
| 3.4 | 5.776 | 21787 | 5.193 | 9795 | 3.793 | 1431 |
| 3.5 | 5.872 | 22151 | 5.281 | 9961 | 3.862 | 1457 |
| 3.6 | 5.967 | 22511 | 5.367 | 10123 | 3.930 | 1482 |
| 3.7 | 6.061 | 22866 | 5.452 | 10283 | 3.997 | 1508 |
| 3.8 | 6.154 | 23217 | 5.537 | 10444 | 4.063 | 1533 |
| 3.9 | 6.246 | 23563 | 5.623 | 10606 | 4.129 | 1557 |
| 4.0 | 6.337 | 23906 | 5.707 | 10764 | 4.194 | 1582 |
| 4.1 | 6.427 | 24245 | 5.788 | 10917 | 4.258 | 1606 |
| 4.2 | 6.516 | 24581 | 5.868 | 11067 | 4.321 | 1630 |
| 4.3 | 6.604 | 24913 | 5.949 | 11221 | 4.384 | 1654 |
| 4.4 | 6.691 | 25241 | 6.031 | 11375 | 4.447 | 1677 |
| 4.5 | 6.777 | 25566 | 6.109 | 11522 | 4.508 | 1701 |
| 4.6 | 6.863 | 25888 | 6.184 | 11665 | 4.570 | 1724 |
| 4.7 | 6.947 | 26207 | 6.265 | 11817 | 4.630 | 1747 |
| 4.8 | 7.031 | 26523 | 6.340 | 11958 | 4.690 | 1769 |
| 4.9 | 7.114 | 26836 | 6.416 | 12102 | 4.750 | 1792 |
| 5.0 | 7.196 | 27146 | 6.492 | 12244 | 4.809 | 1814 |
| 10 | 10.647 | 40163 | 9.659 | 18219 | 7.312 | 2758 |
| 20 | 15.613 | 58897 | 14.236 | 26852 | 10.974 | 4140 |

**Table S3** Variation of lava flow velocity and relative Reynolds number for lavas flows with thickness of 1 m, 5 and 10 m as a function of viscosity (Pa s), topographic slope (°), and density (Kg/m3).

**Table S4**

| flow velocity | Effusion rate | Stephan-Boltzmann constant<br>σ | emissivity<br>ε | Thot<br>temperature of exposed molten lava | Tcrust<br>Temperature of the crust |
|---|---|---|---|---|---|
| $m\ s^{-1}$ | $m^3 s^{-1}$ | $W\ m^{-2}\ K^{-4}$ | | K | K |
| 0.7 | 100 | 5.67E-08 | 0.85 | 1623 | 700 |
| 7.2 | 1000 | | | | |
| | 30000 | | | | |

| | fcrust | viscosity | thickness | lava density | latent heat cryst |
|---|---|---|---|---|---|
| | | Pa s | m | $Kg\ m^{-3}$ | $J\ Kg^{-1}$ |
| | 0.9*EXP(-0.16*flow velocity) | 13 | 1 | 2452 | 370000 |
| | | | 5 | | |
| | | | 10 | | |

**Table S4:** FLOWGO parameters used to calculate the heat loss for the flowing mercury lava. More details are reported in Harris and Rowland (2001).

For details about the above parameters please refer to Harris and Rowland [2001]

**Figure 1**

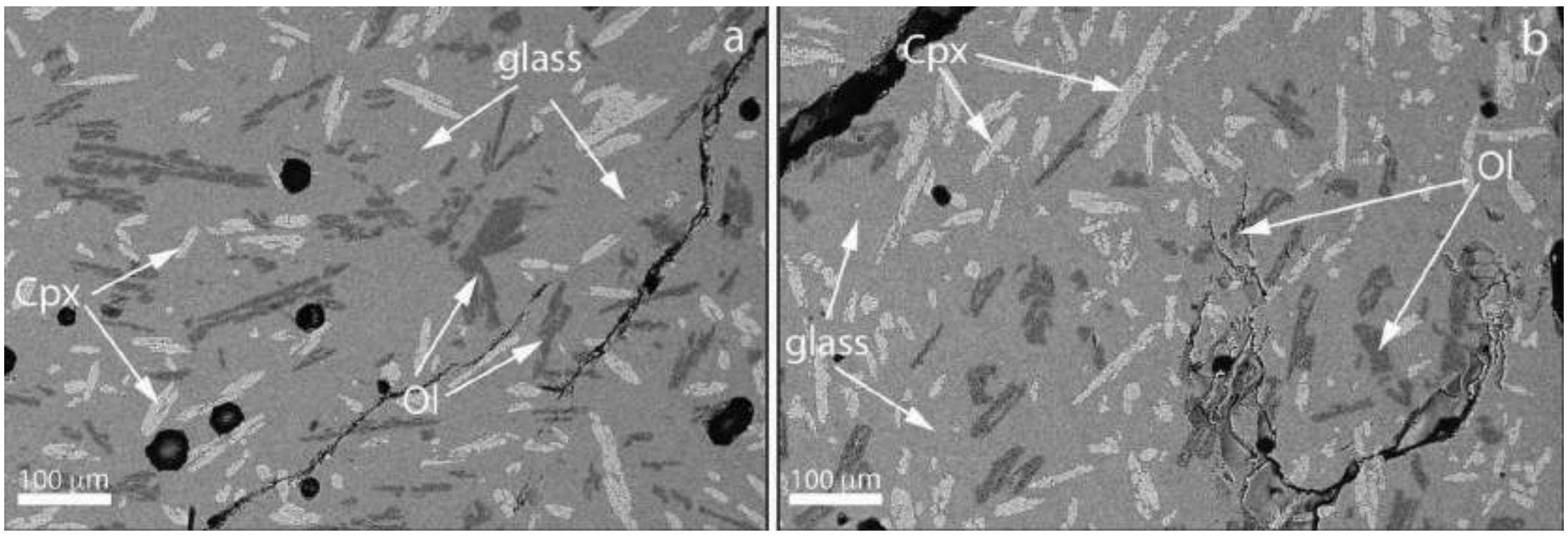

**Figure 1**. Back scattered electron images of representative experimental products. a) Experiment at temperature T=1502 K and shear rate $\dot{\gamma}$=5.0 $s^{-1}$; b) experiment at same temperature as (a) with a shear rate $\dot{\gamma}$=0.1 $s^{-1}$. The two experiments show comparable crystal contents. Labelled crystal phases are olivine (Ol), clinopyroxene (Cpx) and glass.

**Figure 2**

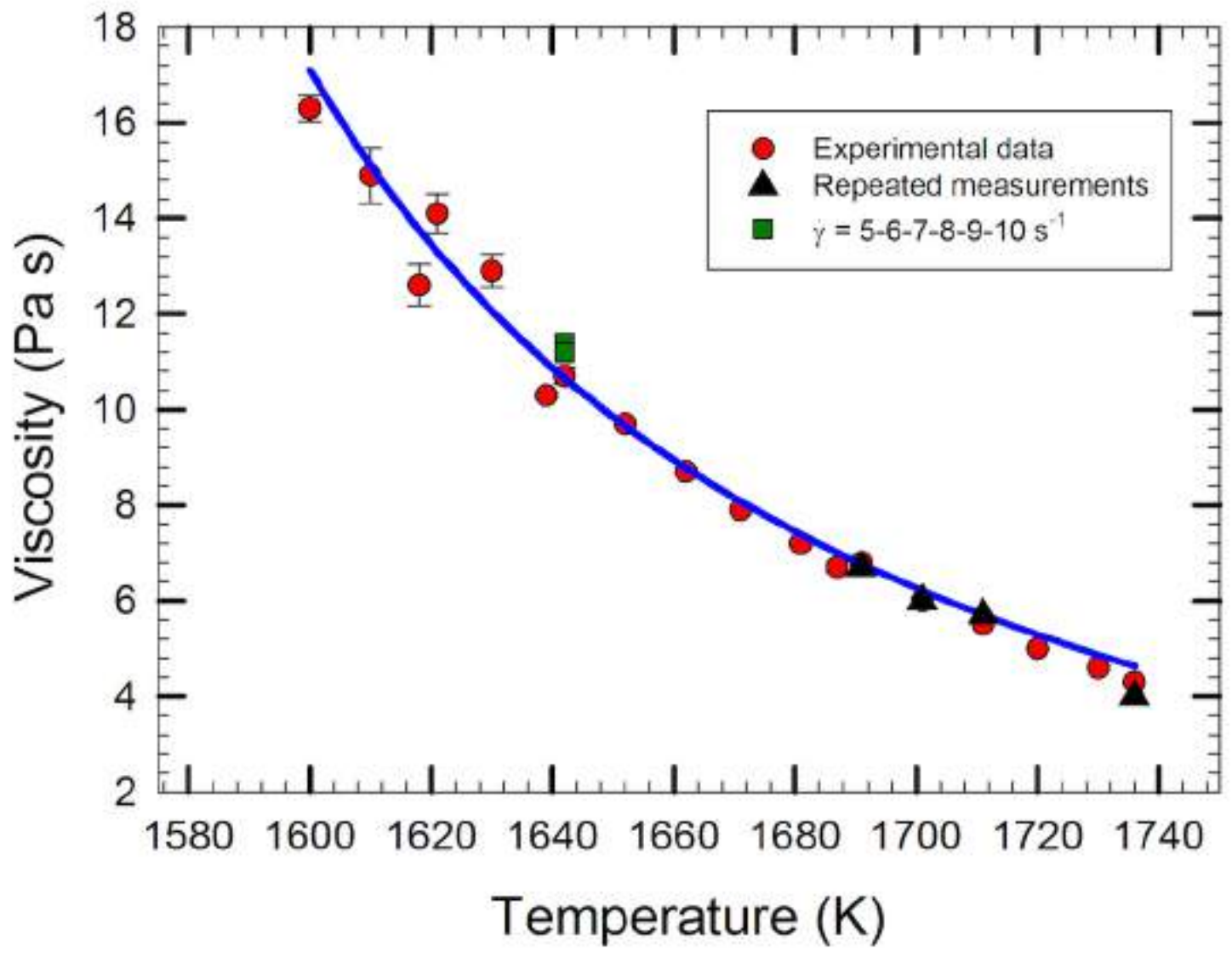

**Figure 2.** Viscosity data for melts at superliquidus conditions. The continuous line represents the predictive model given in Eq. (1). Black full triangles are repeated measurements; green squares represent experiments performed at different shear rates.

**Figure 3**

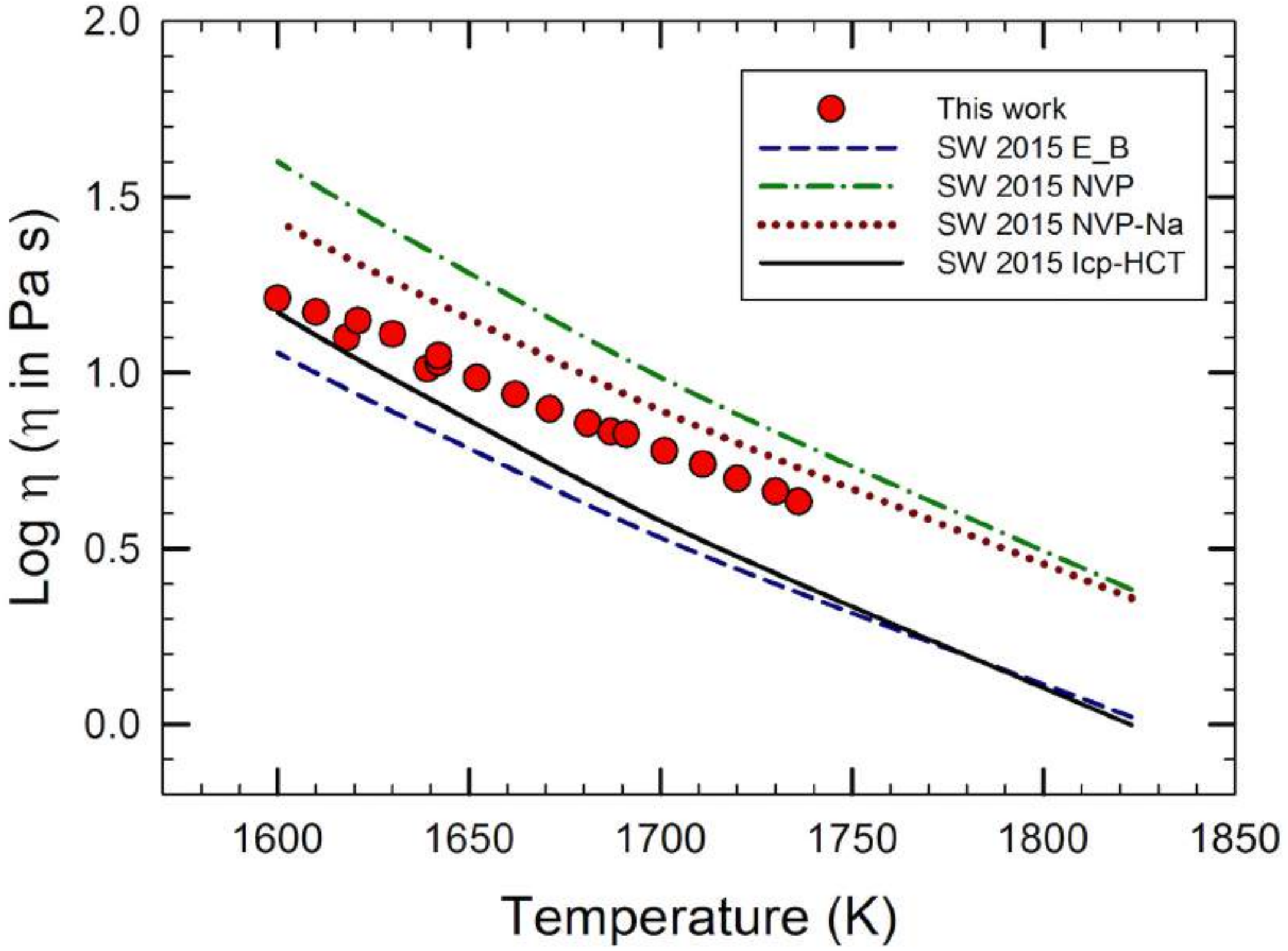

**Figure 3.** Variation of viscosity as a function of temperature showing the comparison between data presented in this work and literature data using possible Mercury compositions. Icp-HCT and E_B refer to basaltic komatiites and Enstatite Basalt, respectively, as reported in *Sehlke and Whittington* [2015]

**Figure 4**

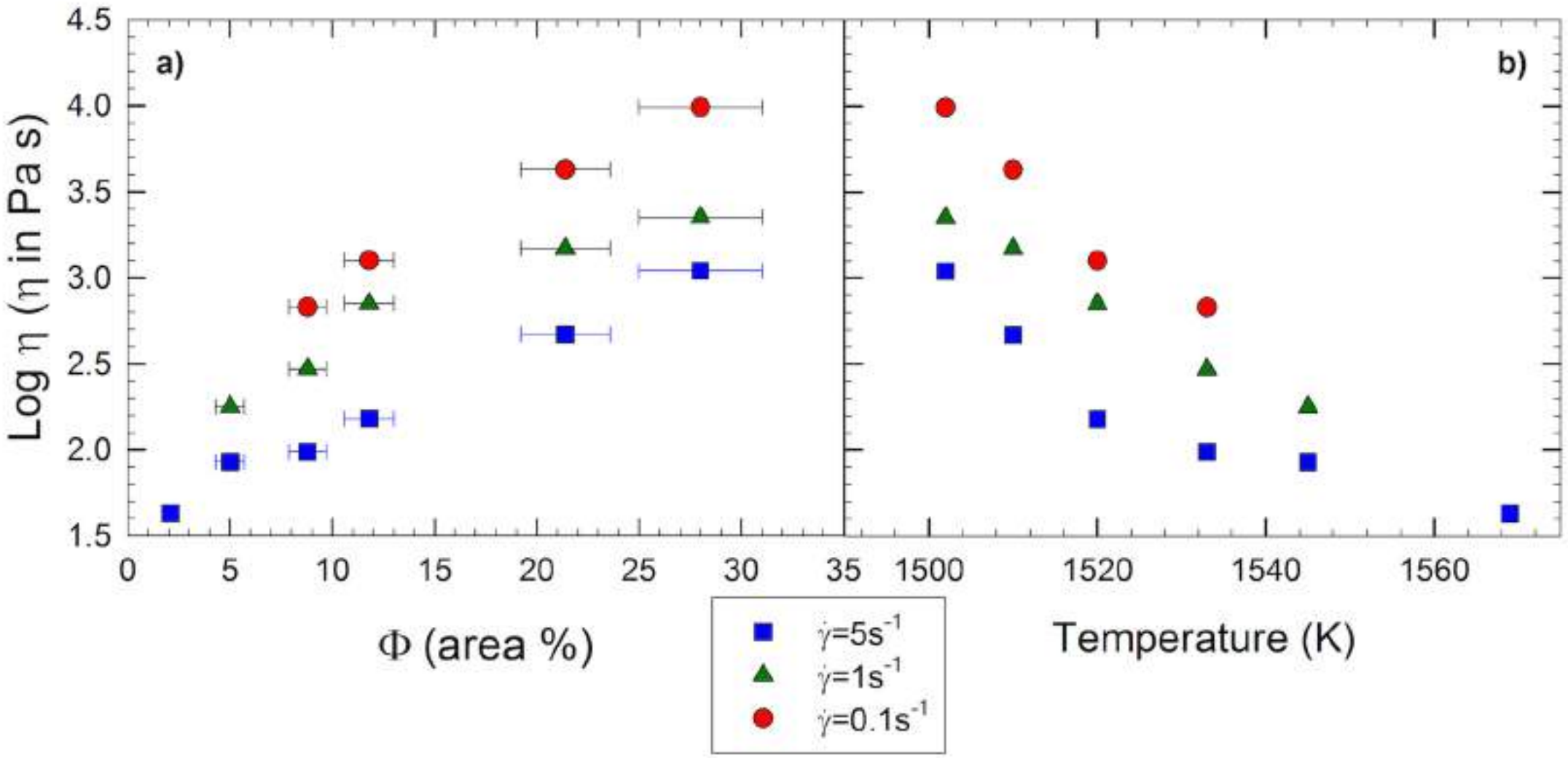

**Figure 4**. **a**) Variation of viscosity [log(η)] as a function of crystal content (Φ area %) at different shear rates ($\dot{\gamma}$); **b**) variation of viscosity [log(η)] as a function of temperature for crystallization experiments at different shear rates ($\dot{\gamma}$). Measurement errors in Figure 4b are comparable or lower than symbol size.

**Figure 5**

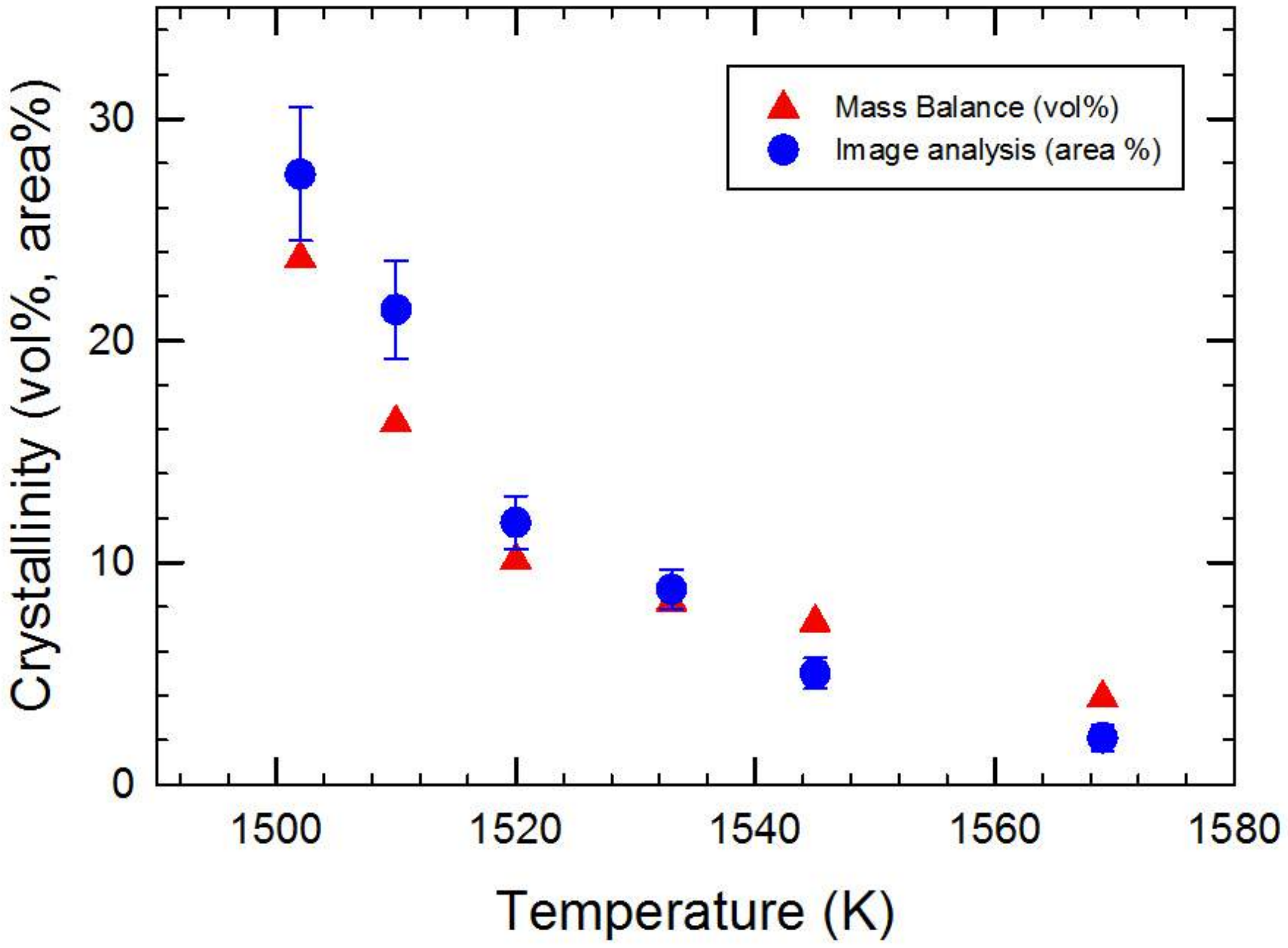

**Figure 5.** Comparison between crystal content obtained by image analysis (area %) and mass balance calculations (vol.%). Errors in area % are on the order of 10 % relative.

**Figure 6**

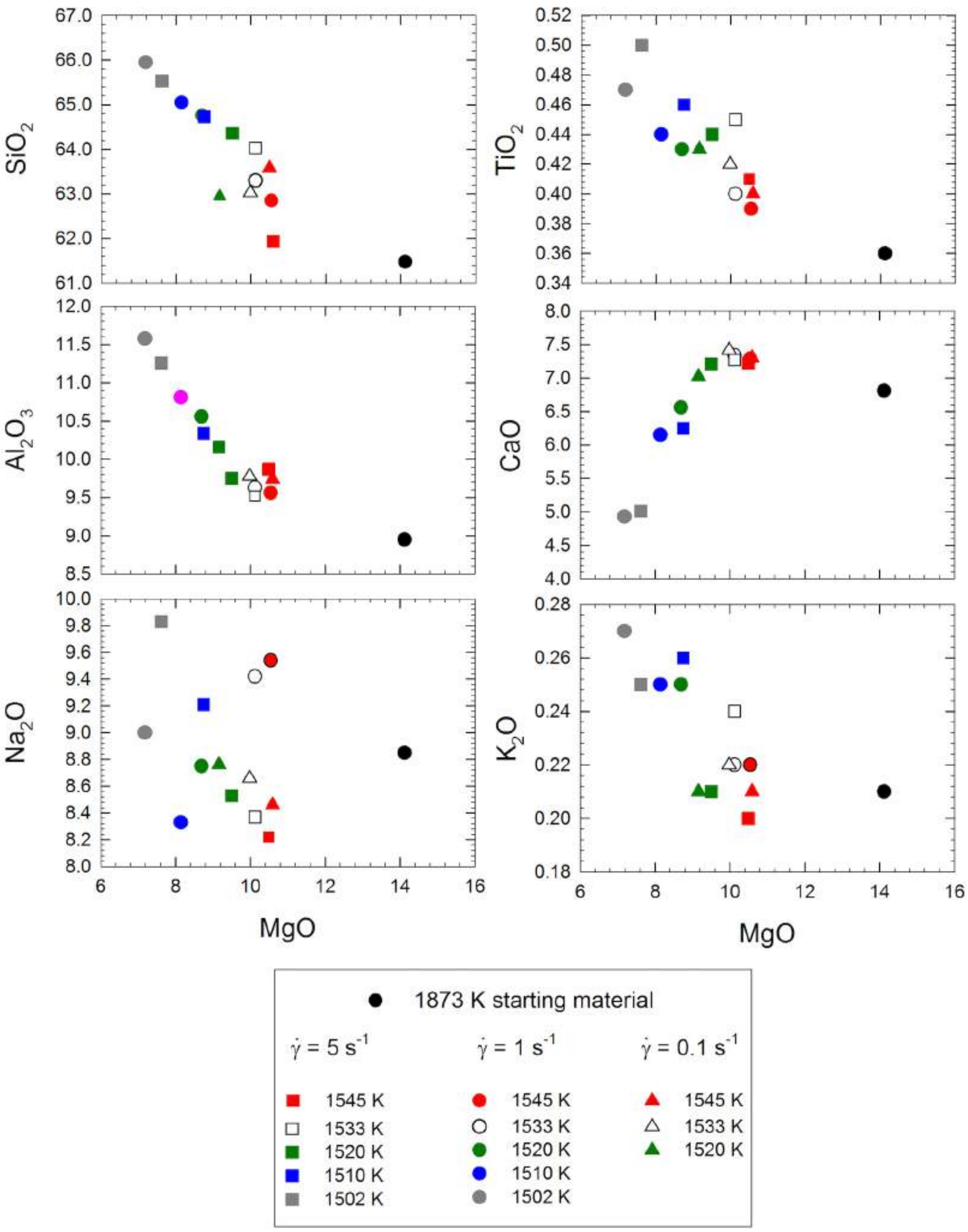

**Figure 6**. Change of the residual glass composition (in wt.% of element concentrations) upon crystallization of the silicate melts. Triangle, circle and square indicate experiments performed at shear rate of 0.1, 1.0 and 5.0 $s^{-1}$, respectively.

**Figure 7**

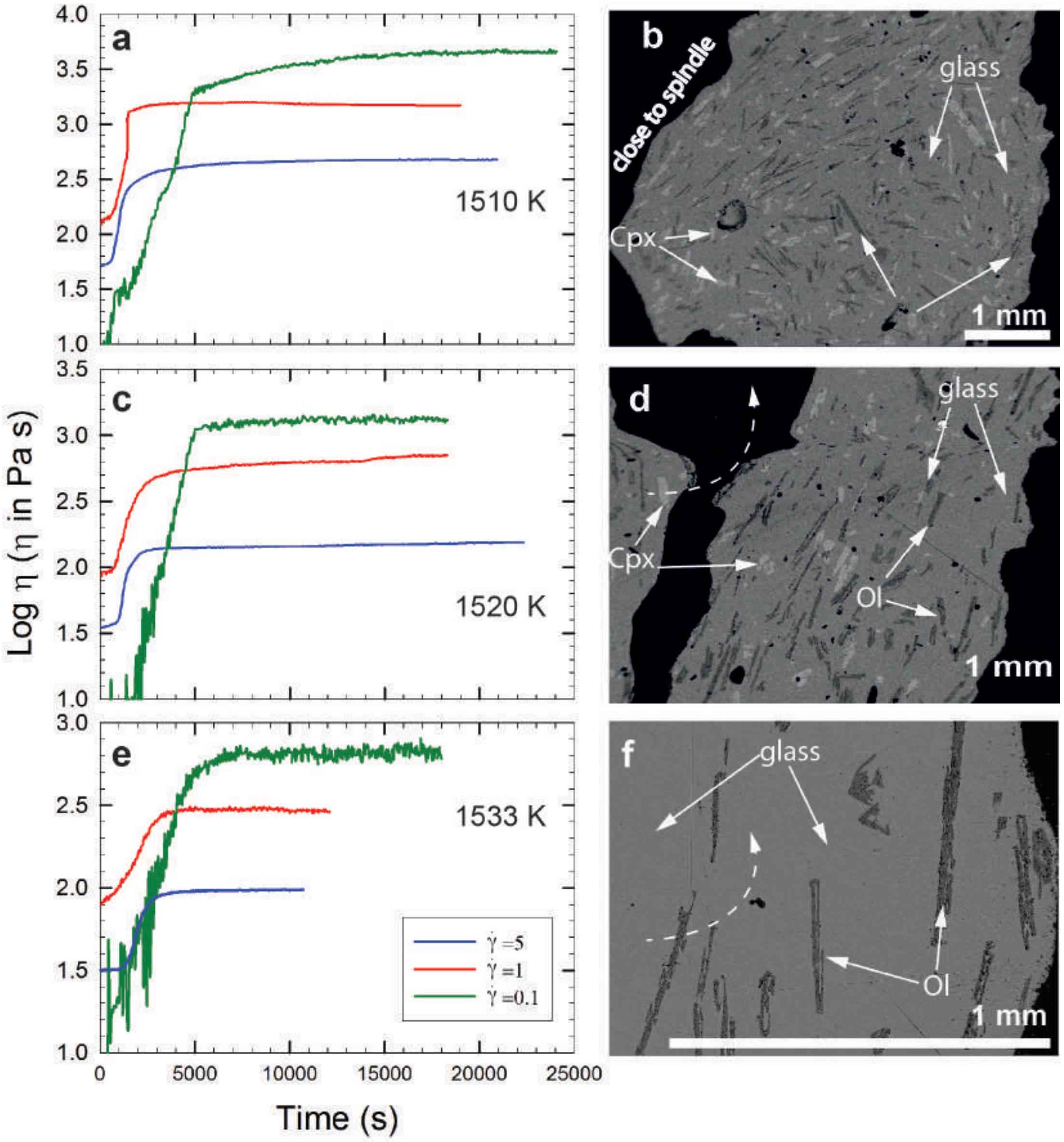

**Figure 7**. **a-c-e**): Variation in viscosity as a function of time and temperature at different shear rates; **b-d-f**): corresponding BSE images and phases after reaching the saturation threshold (flat part of profiles in the plots on the left) for experiments performed at $\dot{\gamma}$ =5 $s^{-1}$. Dashed-lined arrows indicate flow direction; Ol and Cpx refer to olivine and clinopyroxene, respectively.

**Figure 8**

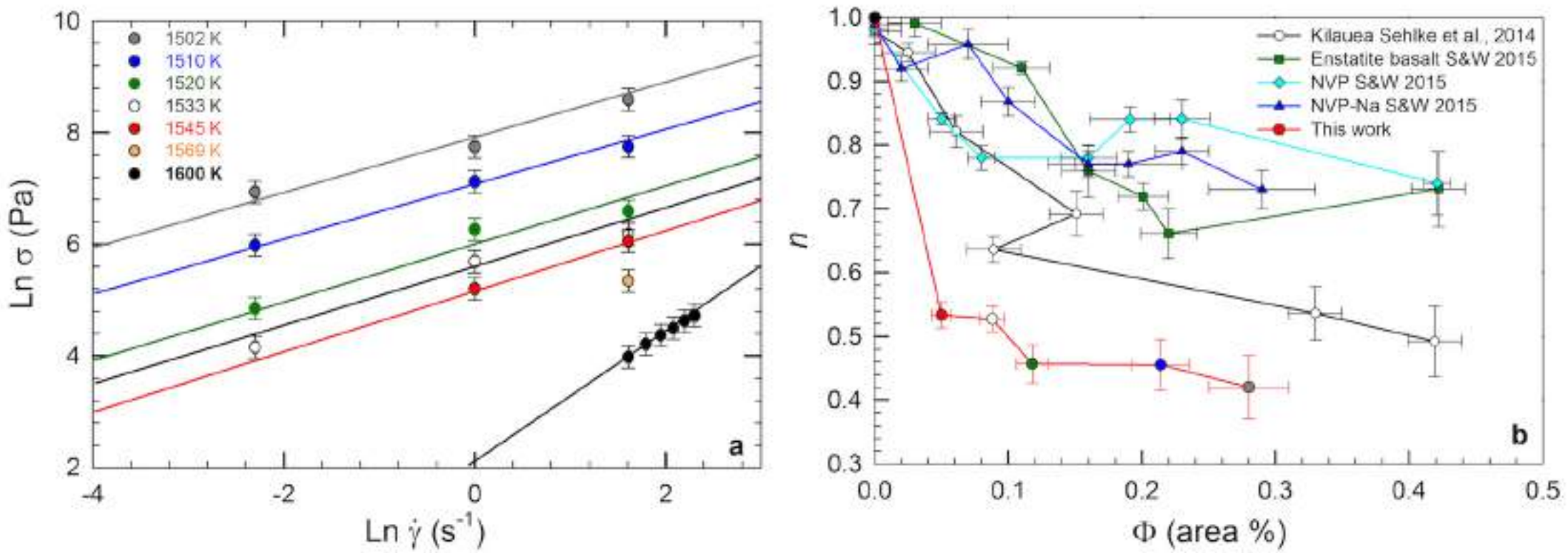

**Figure 8**. **a**) Linear regressions of ln(σ) (shear stress) against ln ($\dot{\gamma}$) (shear rate) for the studied composition. Linear regressions for each dataset are also shown. Note the drastic change in slope when passing from the super-liquidus experiment (1600 K) to the crystal-bearing experiments due to the onset of crystallization; **b**) flow index derived from the linear regression of data in (**a**). Estimated rheological parameters (*n* and *k*) are given in Table 3.

**Figure 9**

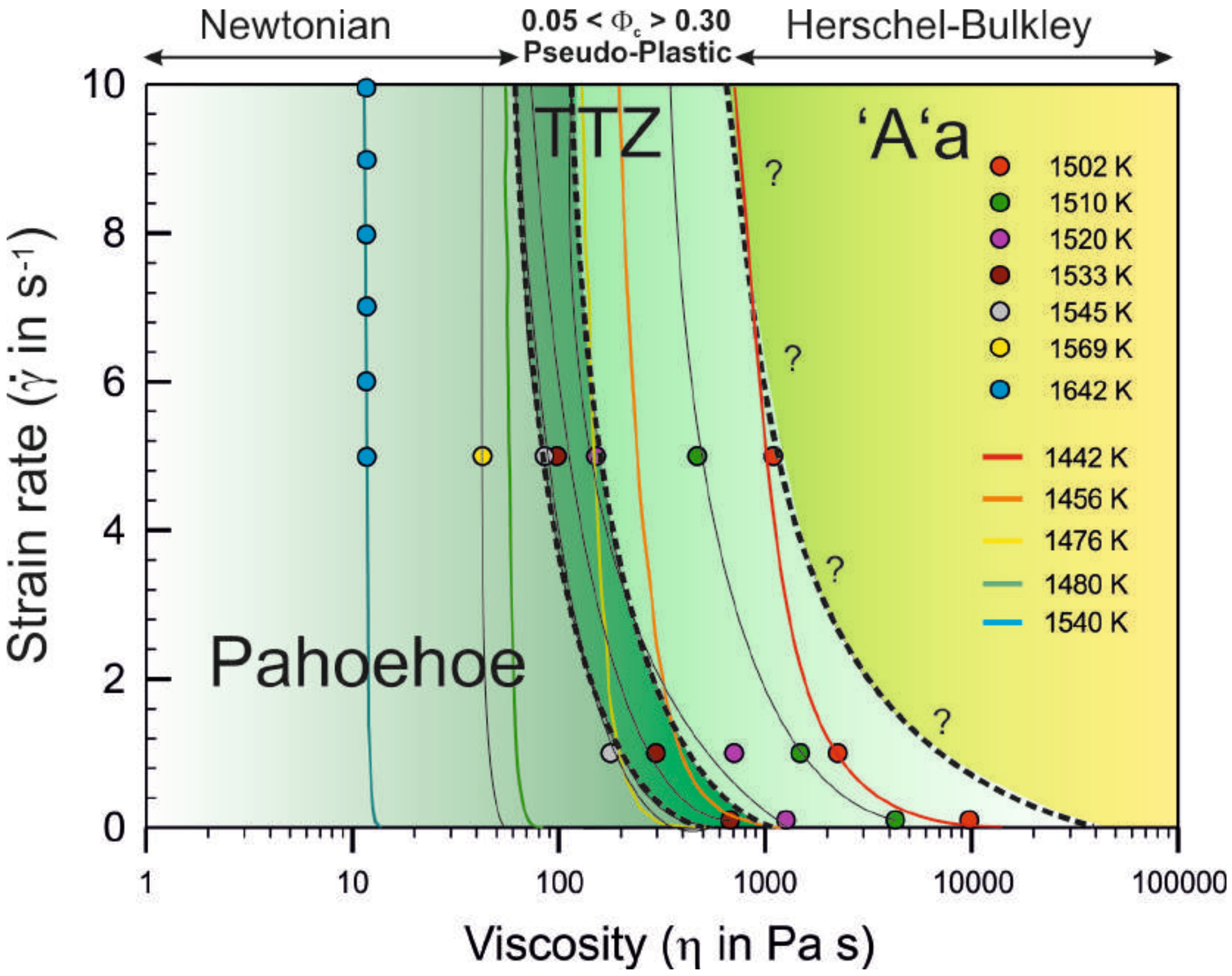

**Figure 9**. Flow curves for the analog Mercury lavas from this work superimposed on the pahoehoe to \`a\`a transition diagram from *Sehlke et al*. [2014]. Coloured lines from right to left refer to data from *Sehlke et al*. [2014] for Hawaiian lavas; dots refer to data from this study. Lava on Mercury start the transition from pahoehoe to \`a\`a at a temperature of ~1533 ± 10 K. The question mark, as reported in *Sehlke and Whittington* [2015], emphasizes the approximate location of the end of the transition threshold zone TTZ.

**Figure 10**

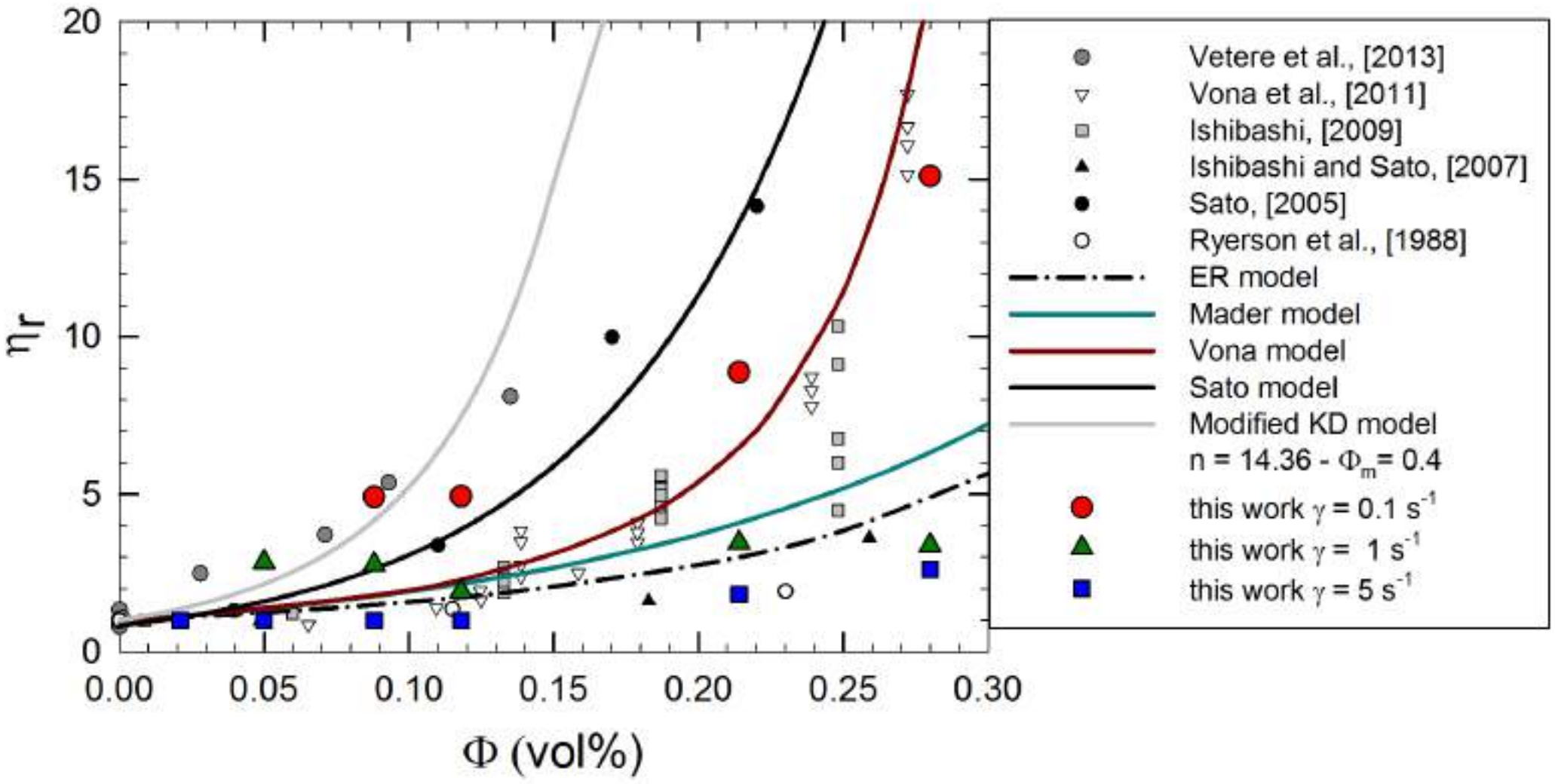

**Figure 10.** Relationship between relative viscosity ($\eta_r$) and crystal fraction ($\Phi$; measured by image analysis, see text for details) for the analog melt in this study (coloured symbols) and literature data. Curves correspond to different models: ER, Einstein-Roscoe model [*Einstein*, 1906; *Roscoe*, 1952]; KD, Krieger-Dougherty model [*Krieger and Dougherty*, 1959]; Sato model [*Sato*, 2005]; Mader model [*Mader et al.*, 2013]; Vona model [*Vona et al.*, 2011] see also supplementary information for details).

**Figure 11**

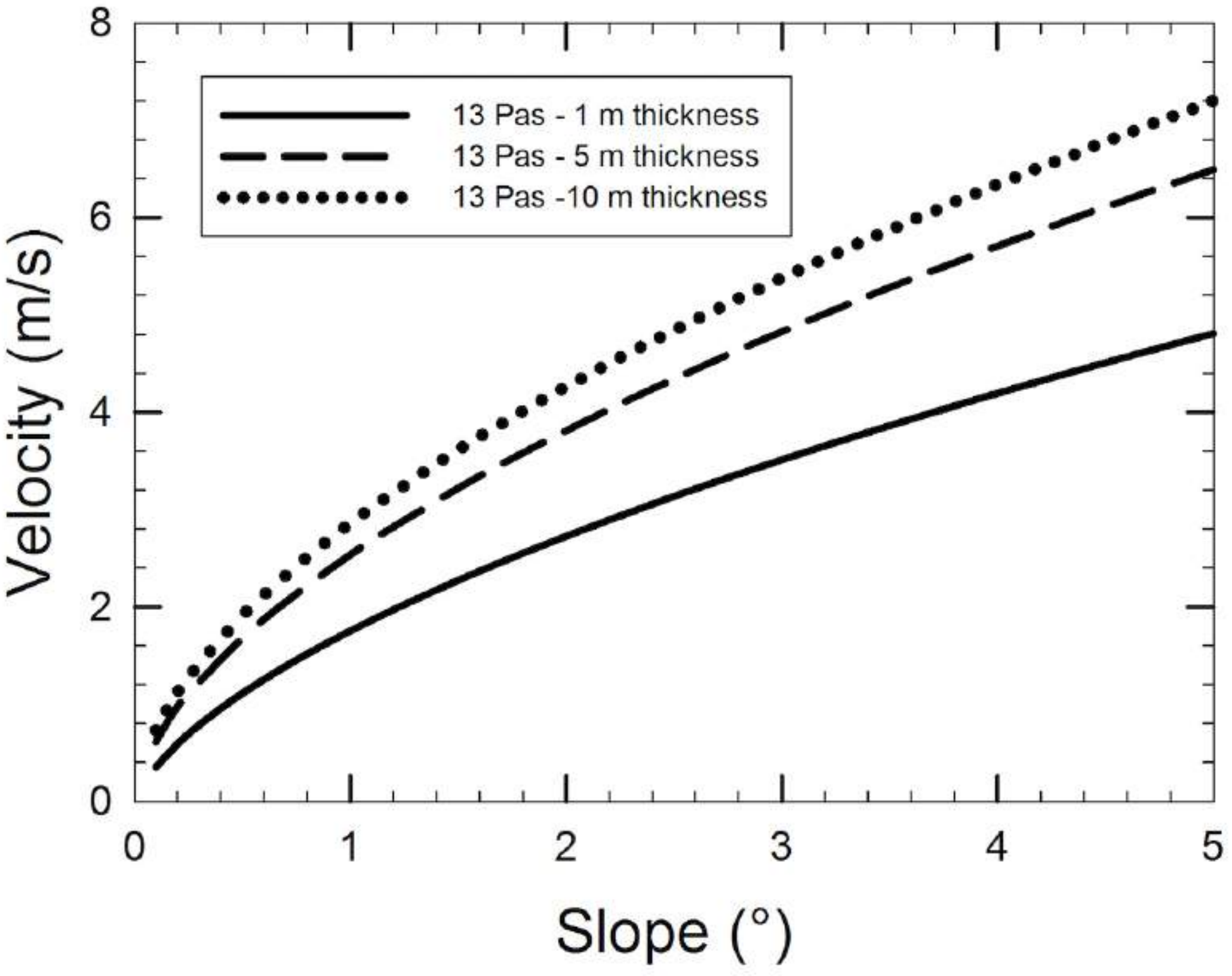

**Figure 11.** Variation of lava flow velocity as a function of topographic slope for three lava flows with thickness of 1 m, 5 m and 10 m (Eq. 2–4).

**Figure 12**

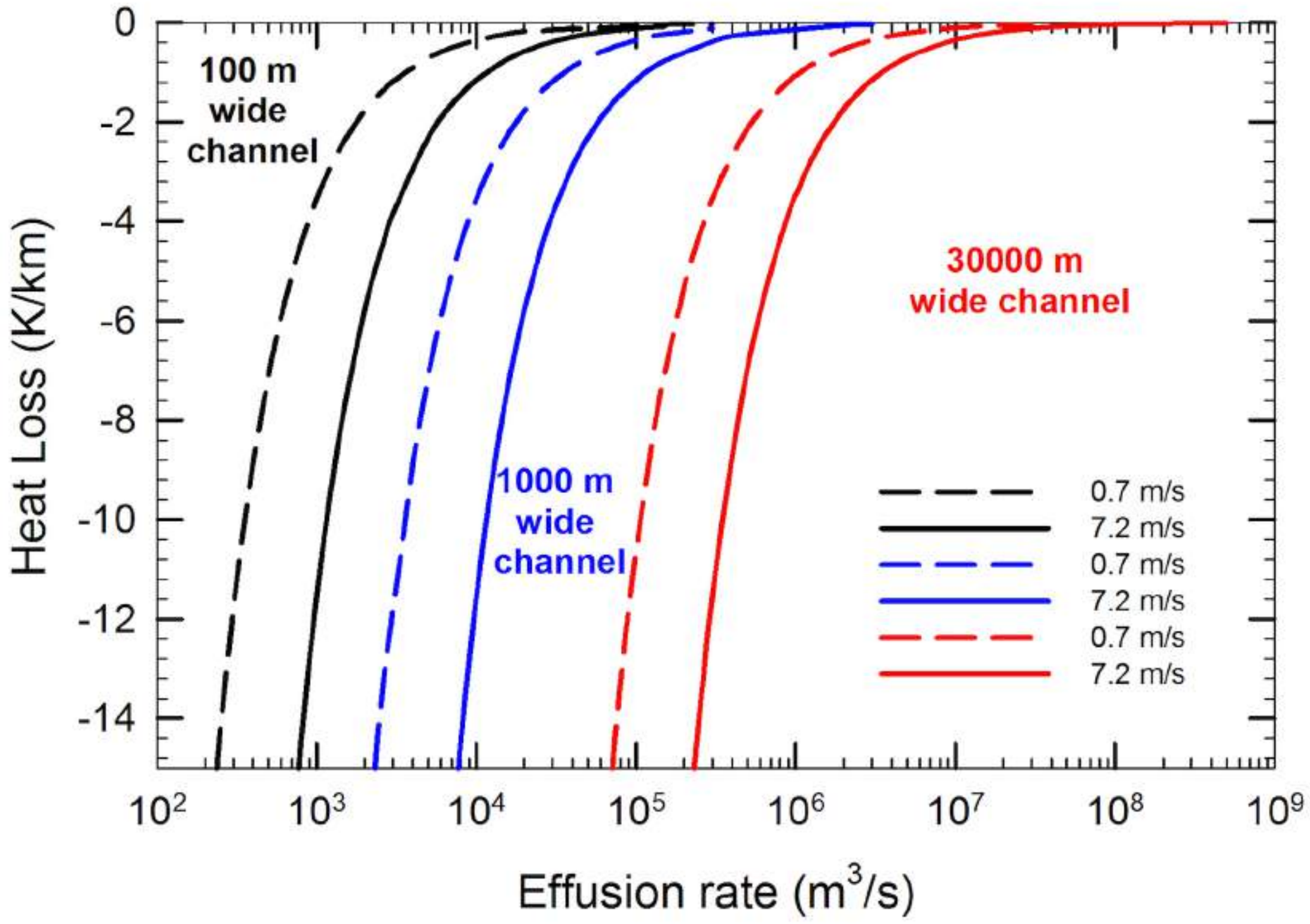

**Figure 12.** Variation of heat loss (in K/km) as a function of effusion rates for a lava flowing in 100, 1000 and 30000 m width channels, with velocities of 0.7 and 7.2 m/s respectively. Details are provided in the main body of text.

**Figure S1**

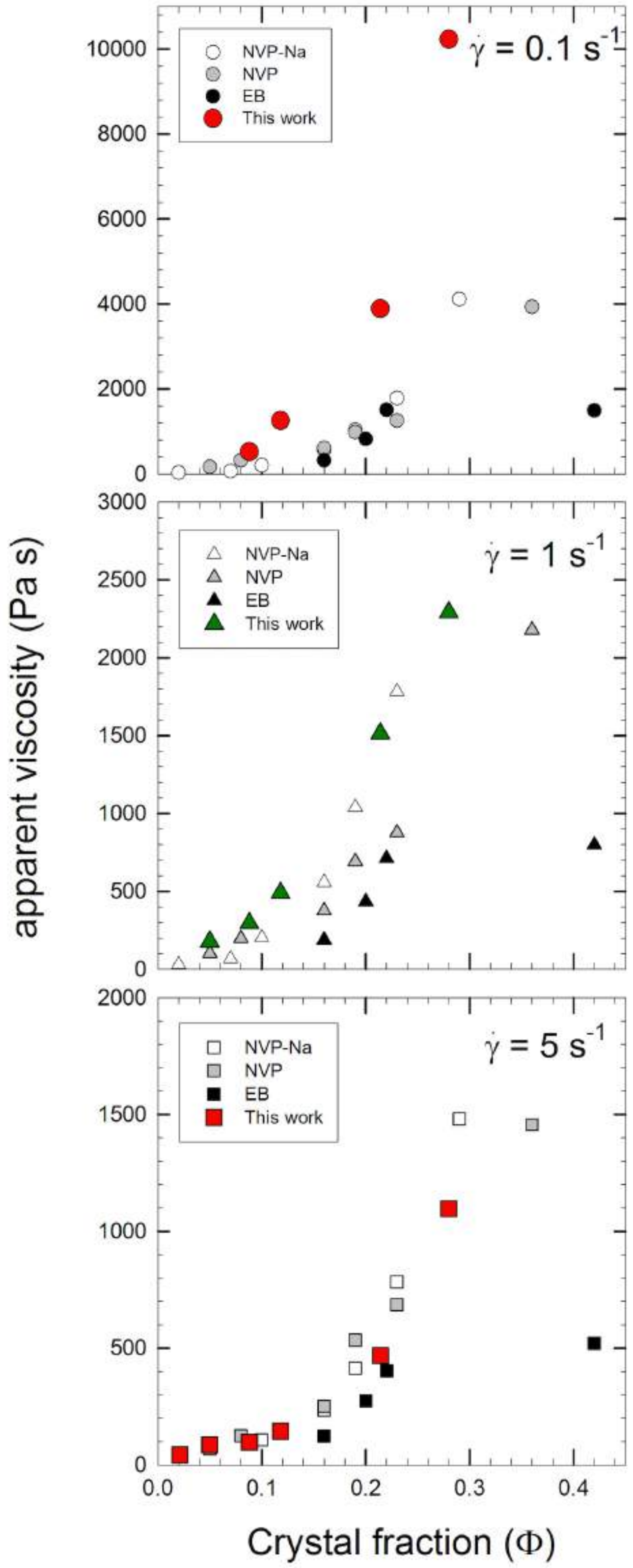

**Figure S1**

Variation of the apparent viscosity as a function of crystal content for NVP-Na, NVP and EB compositions Studied by Sehlke and Whittington [2015] and our new experimental data.